\documentclass[ reprint,twocolumn, amsmath,amssymb, aps,]
{revtex4-1}

\usepackage{graphicx}
\usepackage{dcolumn}
\usepackage{bm}
\usepackage{float}
\usepackage{hyperref}
\usepackage{subfig}
\usepackage{dsfont}
\usepackage{slashed}
\usepackage{feynmp}
\usepackage[section]{placeins}
\usepackage{braket}
\usepackage{upgreek}
\usepackage[normalem]{ulem}

\newcommand{\qqb}{Q\bar{Q}}
\newcommand{\QCD}{\mathrm{QCD}}
\newcommand{\corr}{{\mathrm{corr}}}
\newcommand{\tr}{\mathrm{tr}}

\begin{document}

\preprint{TIFR/TH/19-42}

\title{Quantum evolution of quarkonia with correlated and uncorrelated noise}

\author{Rishi Sharma}
\email{rishi.sharma@gmail.com}
 \affiliation{Tata Institute of Fundamental Research \\
 Colaba, Mumbai, India. 400005
 }
\author{Anurag Tiwari}%
 \email{anurag.tiwari128@gmail.com}
\affiliation{%
 Tata Institute of Fundamental Research \\
 Colaba, Mumbai, India. 400005
}

\date{\today}
\begin{abstract}
In the quark gluon plasma (QGP), it is well known that the evolution of
quarkonia is affected by the screening of the interaction between the quark and
the anti-quark. In addition, exchange of energy and color with the surrounding
medium can be included via the incorporation of noise terms in the evolution
Hamiltonian. For noise correlated locally in time, these dynamics have been
studied in a simple setting by Ref.~\cite{akamatsu20181}. We extend this
calculation by considering non-Abelian dynamics for a three dimensional wavefunction. We also
propose a modification of the noise correlation, allowing it to have a finite 
correlation in time with the motivation to include long-lived gluonic 
correlations. We find that in both cases the results differ significantly 
from solutions of rate equations.
\end{abstract}

\maketitle

\section{\label{sec:introduction}Introduction}
The propagation of quarkonia in the QGP is influenced by
several processes. Screening in the thermal medium weakens the interaction
between the $Q$ and the $\bar{Q}$~\cite{matsui19861}. Interaction with ``on
shell'' thermal gluons can lead to dissociation (gluo-dissociation)~\cite{peskin19791,bhanot19791}.
In systems with high occupation numbers of heavy quarks (for example in heavy
ion runs at the LHC) recombination~\cite{Grandchamp:2001pf,grandchamp20021} of
$c$ and $\bar{c}$ may also play an important role. All these effects play a
role in the determination of the experimental observable, $R_{AA}$, which is the
normalized (per binary collision) ratio of the observed quarkonium yields in
heavy ion ($AA$) collisions versus the yields in $pp$ collisions. 

The large mass of the heavy quark, $M$, provides a natural starting point for
the analysis of these effects. There is a clear separation of energy scales
between the mass $M$ of the heavy quark ($\sim 1.5$GeV for $c$ and $4.5$GeV for
$b$) and the scales $\Lambda_{\QCD}$ and the temperature $T\lesssim 500$MeV. In
contrast with open heavy flavors, quarkonia are non-relativistic bound states
and have additional scales: the inverse of the size, $1/r$, and the binding
energy $E_b$. If the strong coupling $\alpha=g^2/(4\pi)$ at the scale $1/r$ is
sufficiently smaller than $1$, then the bound states are Coulombic and these
additional scales can be written in terms of the velocity $v\sim\alpha$:
$1/r\sim Mv$, $E_b\sim Mv^2$. In this case the hierarchy of scales can be
written as $M\gg Mv \gg Mv^2$~\cite{bodwin19941}.

Even with optimistic estimates of $\alpha$, the approximation $\alpha\ll 1$ is not
expected to be quantitatively reliable for most quarkonium states except for
perhaps the lowest $b\bar{b}$ bound state. (One way to see it is that matching
the observed quarkonium spectra requires a long distance piece in the $\qqb$
potential in addition to the Coulombic piece~\cite{eichten19941,eichten20071}). It is 
assumed more generally that a non-relativistic treatment of quarkonia is still
valid with the hierarchy $M\gg 1/r\gg E_b$.

This hierarchy in scales allows for application of an effective field theory (EFT)
treatment of the system which is valid at the lowest energy scale $E_b$. At the
lowest order in $rE_b$, the EFT consists of non-relativistic quarks bound by a
potential~\cite{pineda19981} (see Ref.~\cite{brambilla20001} for a
comprehensive review). At higher order the theory
features interactions mediated by
gluons of wavelength $1/E_b$. Effects of higher order terms are suppressed by positive
powers of $rE_b$, where factors of $r$ can be seen as arising from a long wavelength
expansion of the fields. This framework is called pNRQCD. 

At $T=0$ the potential can be calculated
using non-perturbative techniques (\cite{stack19841}).
At finite $T$, the coupling between $Q$ (and $\bar{Q}$) and the gluons in the
thermal medium at the energy scale $T$, and the coupling between the medium
gluons at that scale, also play a role. It is typically assumed that $1/r\gg
T,\Lambda_{\QCD}$ but the relative hierarchy between $E_b$, $T$, and
$\Lambda_{\QCD}$ is unclear. A finite temperature version of
pNRQCD~\cite{brambilla20081} has been developed to analyze this system

It is well known that the QGP medium formed in heavy-ion collisions such as
RHIC and LHC is best described as a strongly-coupled medium. Therefore, the ultimate goal should be to use EFT
methods to write observables in terms of quantities which can be calculated on lattice. As a concrete
example, the singlet potential has been computed on the
lattice~\cite{kaczmarek20031,rothkopf20111,Bala:2019cqu}. 

However, non-perturbative calculations of some relevant dynamical processes is
still challenging and weak-coupling calculations are still useful. An important
result in weak-coupling was obtained in Ref.~\cite{laine20071} which showed
that the Wilson loop of heavy quarks which is related to the potential between
quark-antiquark pair, is complex at finite $T$. Furthermore, it was
shown~\cite{brambilla20081} that pNRQCD naturally incorporates the process
known as gluo-dissociation (\cite{peskin19791,bhanot19791}) as its dynamical
degree of freedom include low energy gluonic degrees of freedom (and other
light degrees of freedom if any) in addition to the wavefunctions of $\qqb$
pair. 

Such weak-coupling calculations have given insight into the problem
and results from these calculations can be used to obtain semi-quantitative
estimates for experimental observables of interest: for example $R_{AA}$ in
heavy ion collisions.

Many such calculations have attempted to address the phenomenology of
quarkonium states in the QGP. For approaches using a medium modified $T$-matrix
approach see
Refs.~\cite{grandchamp20041,rapp20091,zhao20111,emerick20111,zhao20121,du20171,du20181}.
For approaches based on gluo-dissociation see
Refs.~\cite{brezinski20121,nendzig20131,hong20191}. For approaches based on
the complex potentials derived by~\cite{laine20071} see
Refs.~\cite{laine20071,strickland20111,strickland20112,margotta20111,krouppa20151,krouppa20171,krouppa20181}.
For approaches including recombination see
Refs.~\cite{Grandchamp:2001pf,grandchamp20021,greco20031,zhang20021,capella20071,bravina20081,yan20061,peng20101,borghini20111,song20111,ferreiro20121,blaizot20151} (See
~\cite{andronic20111,kostyuk20031,thews20061,gupta20141} and
references therein for statistical approaches). For quarkonia at high $p_T$ see
~\cite{sharma20121,aronson20171,makris20191}. For approaches based on Schr\"{o}dinger-Langevin equation see
Refs.~\cite{Katz:2015qja,Gossiaux:2016htk,Bernard:2016spw}. For a comprehensive
review of the phenomenology of heavy quarks and quarkonia see
Ref.~\cite{andronic20151} and references therein. 

In the remaining part of this introduction we will review aspects of the theory
particularly relevant for our work to set up our calculation.

\subsection{\label{subsec:decoherence11} Theory Overview}

In Ref.~\cite{laine20071} the $\qqb$ system was analyzed in weak-coupling
in the regime where the relevant energy scales satisfy the hierarchy $E_b\ll
1/r \ll T$. With these assumptions it was proved that at late times, the time
evolution equation for a thermal averaged correlator for a static $\qqb$
pair, $\langle\Psi_{\qqb}(\vec{r},t)\Psi_{\qqb}(\vec{0},0) \rangle$ satisfies
a Schr\"{o}dinger like equation. The evolution kernel has an imaginary piece
with the formal structure of an imaginary potential which arises due to the
Landau damping of the gluons exchanged between the $Q$ and the $\bar{Q}$ due to
thermal gluons. 

From the complex potential one can calculate the thermal width of quarkonia in
the medium. Interpreting the inverse width for a quarkonium state as its decay
rate one can solve the rate equation to find the fraction
of quarkonia that survive in the medium during its evolution. Thus one has a
theoretical calculation for $R_{AA}$ for various quarkonium
states~\cite{laine20071,strickland20111,strickland20112,margotta20111,krouppa20151,krouppa20171,krouppa20181}.

In Refs.~\cite{brambilla20081,brambilla20101,brambilla20111,brambilla20131} the calculation was extended by
considering different hierarchies of the energy scales (between $1/r$, $E_b$, $T$, 
$\Lambda_{\QCD}$), and additional processes like gluo-dissociation, within the weak coupling
approximation using pNRQCD. Boltzmann equations in weak coupling have been
written down and solved in
Refs.~\cite{blaizot20151,blaizot20171,blaizot20181}. In 
Refs.~\cite{yao20171,yao20184,yao20183}, a Lindblad equation was derived and
used to obtain a Boltzmann transport equation and compute $R_{AA}$.  

However most calculations of $\qqb$ described above ignore the coherence of the quarkonium
wavefunction on the time scale of the medium evolution. Therefore one requires
a formalism which tracks the full quantum evolution of the $\qqb$ state.

The correct way to dynamically interpret the results obtained in~\cite{laine20071} is to look at the evolution of the $\qqb$ density matrix by treating the $\qqb$ pair as an open quantum system~\cite{akamatsu20131,akamatsu20151}. The complex potential corresponds to the
decoherence of a $\qqb$ state. In addition, another process --- dissipation (which is required for heavy-quarks equilibration but is expected to be small for tightly bound
quarkonia~\cite{akamatsu20131,akamatsu20151}, however see Ref.~\cite{akamatsu20191}) can also be naturally derived in this formalism~\cite{akamatsu20151}. This approach to quarkonium dynamics was introduced in various physical regimes in Refs.~\cite{akamatsu20121,borghini20121,young20131}. It was developed in the weak coupling regime in Refs.~\cite{akamatsu20121,akamatsu20131,akamatsu20151,akamatsu20181}, in the pNRQCD framework in Refs.~\cite{brambilla20171,brambilla20181}, and more recently in Ref.~\cite{yao20191}.

\subsection{\label{subsec:decoherence12} Summary}
In this paper we follow the formalism developed in
Refs.~\cite{akamatsu20131,akamatsu20151}. In the weak coupling regime the
authors derived equations for the evolution of the density matrix for the
$\qqb$ system in contact with a thermal medium. It undergoes decoherence, which
refers to processes where interactions with the environment convert a pure
quantum state of the system to a mixed state. In this context it refers to
scatterings with the medium gluons. If the typical energy scale of the system
(here $E_b$, which is inverse of the system time scale) is much smaller than
the environment relaxation rate, then the system evolution during a typical
interaction can be taken to be slow. Formally taking the system frequency to be
much smaller than $gT$, in
Refs.~\cite{laine20071,akamatsu20151} a Markovian master equation in Lindblad
form was derived. Then, the evolution is
only controlled by two parameters - the temperature $T$ and value of the
strong coupling $g$. These evolution equations can be naturally solved by
introducing appropriate noise fields, solving the resulting stochastic
Schr\"{o}dinger equations, and taking the ensemble average~\cite{gisin19921}.
Refs.~\cite{laine20071,akamatsu20151} derived the corresponding stochastic
Schr\"{o}dinger equation with a noise term which is correlated locally in time.  

In Ref.~\cite{akamatsu20181}, the authors solved a simplified version of these
equations for one dimensional wavefunctions and ignoring the color structure. We expand their
implementation into a more general setup with a simplification which we argue
from the viewpoint of pNRQCD. We implement stochastic Schr\"{o}dinger equations
which keep track of the color, angular-momentum and radial wavefunction in
position space for the quarkonia pair. This is the main technical advance
presented in our paper.

In Sec.~\ref{sec:pNRQCD}, we propose a modification to the stochastic
Schr\"{o}dinger equation which can incorporate finite-frequency processes. We
argue that the process of absorption or emission can be described if the noise
field is allowed to be correlated in time with a finite time scale. This makes
the system evolution non-Markovian due to memory-effects in the bath degrees of
freedom. The modification can be checked by comparing the results at early time
with classical decay approach.   

The brief outline of the paper is as follows. In Sec.~\ref{sec:decoherence1}, we extend the calculation of
\citep{akamatsu20181} to a realistic three dimensional case while keeping the
complete color structure of the $\qqb$ pair. We also make an expansion in
small $\vec{r}$ for the noise fields. A small $\vec{r}$ expansion is justified
as long as $\langle r \rangle m_D \ll 1$.  In Sec.~\ref{sec:pNRQCD} we
extend the stochastic equation used in Sec.~\ref{sec:decoherence1}, by allowing
the noise fields to be correlated in time. This allows us to perform a
quantum calculation of gluo-dissociation.
Our main results for the above two cases are presented in Sec.~\ref{sec:final_results}. We
also make a comparison with simple rate-equation like approaches which has been
traditionally used in phenomenological approaches. 
 
Finally, in the appendix we provide a comparison between a $\vec{r}$ expanded
and a calculation without making the expansion (``un-expanded'') for a simple one-dimensional colorless system,
for which results were available from \cite{akamatsu20181}.
\section{\label{sec:decoherence1}Decoherence in small $\vec{r}$ limit}
In this section, we briefly review the evolution equations for quarkonia in the
QGP~\cite{akamatsu20131,akamatsu20151} and simplify them using the
approximation $r\ll 1/T$.
\subsection{\label{subsec:akamatsu2015} Master equation for the quarkonium density matrix 
$\rho_{\qqb}(t)$}
The $\qqb$ ``system'' continuously exchanges energy with the thermal
``environment''.  The density matrix ($\rho$) of the $\qqb$ is obtained by tracing out
the environmental degrees of freedom. In general the process of tracing out the
environmental degrees is complicated. However tractable evolution equations for
the system can be obtained under some simplifying equations. The starting point
of our calculation is the evolution equation for the $\qqb$ density matrix
(Eq.~\ref{eqn:aka_reduced_density}) derived in
Refs.~\cite{akamatsu20131,akamatsu20151} using the following approximations.

\begin{enumerate}
\item{All interactions are governed by a single coupling constant $g$
and it was assumed that $g\ll1$  The evolution equation was derived keeping
terms up to $\mathcal{O}(g^2)$}
\item{It was assumed that $E_b\ll gT$. Physically this corresponds to assuming that the thermal
gluons relax [on a rough time scale $\sim 1/(gT)$] on a shorter time scale than the natural
time scale for the system oscillations [$\sim 1/(E_b)$]. Then each exchange with the
environment can be treated as independent and hence the density matrix
evolution is Markovian: the operator governing the evolution of $\rho$ does not
depend on the history and is local in time. Given that the scales 
$T$ and $E_b$ are not well separated, it is worth scrutinizing this assumption
further, and we will do this in Sec.~\ref{sec:pNRQCD}.}
\item{It was assumed that $M$ is much greater than
any other scale in the system. Then the Hamiltonian for the fermionic part can
be expanded in powers of $1/M$\cite{foldy19501}. Only the leading
order terms in $1/M$ were retained.}
\item{Under the further assumption that $M \gg T$, dissipation terms are
smaller than terms leading to the decoherence of the wavefunction which is the
regime that we will focus on here.}
\end{enumerate}

Using these approximations, a master-equation for the $\qqb$ pair was derived
in Lindblad form \cite{akamatsu20151,lindblad19761}
\begin{widetext}
\begin{equation}
\label{eqn:aka_reduced_density}
    \frac{\partial}{\partial t}\left(\begin{array}{c}{\rho_{1}} \\ {\rho_{8}}\end{array}\right)_{(t, \vec{r}, \overline{s})}  =\left(i \frac{\vec{\nabla}_{r}^{2}-\vec{\nabla}_{s}^{2}}{M}\right)\left(\begin{array}{c}{\rho_{1}} \\ {\rho_{8}}\end{array}\right)_{(t, \vec{r})}+i(V(\vec{r})-V(\vec{s}))\left[\begin{array}{cc}{C_{\mathrm{F}}} & {0} \\ {0} & {-1 / 2 N_{\mathrm{c}}}\end{array}\right]\left(\begin{array}{c}{\rho_{1}} \\ {\rho_{8}}\end{array}\right)_{(t, \vec{r}, \vec{s})}+\mathcal{D}(\vec{r}, \vec{s})\left(\begin{array}{c}{\rho_{1}} \\ {\rho_{8}}\end{array}\right)_{(t, \vec{r}, \vec{s})}.
\end{equation}{}
Here $\vec{r}$ corresponds to the relative separation between the $\qqb$ in the
``ket'' space and $\vec{s}$ is the separation in the ``bra'' space.  $M/2$ is
the reduced mass of the $\qqb$ system.
$\rho_{1,\;8}=\rho_{1,\;8}(t,\vec{r},\vec{s})$ are the singlet and octet
components of the $\qqb$ density matrix in position space. $V(\vec{r}),
V(\vec{s})$ correspond to the potential between $Q$ and $\bar{Q}$.
$N_{\mathrm{c}}$ is the number of color degree of freedom and $C_{F} =
(N_{\mathrm{c}}^2-1)/2N_{\mathrm{c}}$. We consider the $\qqb$ pair at rest in
the medium and hence the center-of-mass coordinates $\vec{R}$, $\vec{S}$ do not
play a role and we have suppressed the dependence on them.

$\mathcal{D}(\vec{r},\vec{s})$ are terms related to  decoherence of the $\qqb$
state~\cite{akamatsu20151},

\begin{eqnarray}
\label{eqn:aka_decoherence_terms}
&& \mathcal{D}(\vec{r}, \vec{s})=2 C_{\mathrm{F}} D(\overrightarrow{0})-(D(\vec{r})+D(\vec{s}))\left[\begin{array}{cc}{C_{\mathrm{F}}} & {0} \\ {0} & {-1 / 2 N_{\mathrm{c}}}\end{array}\right]-2 D\left(\frac{\vec{r}-\vec{s}}{2}\right)\left[\begin{array}{cc}{0} & {1 / 2 N_{\mathrm{c}}} \\ {C_{\mathrm{F}}} & {C_{\mathrm{F}}-1 / 2 N_{\mathrm{c}}}\end{array}\right]+2 D\left(\frac{\vec{r}+\vec{s}}{2}\right)\left[\begin{array}{cc}{0} & {1 / 2 N_{\mathrm{c}}} \\ {C_{\mathrm{F}}} & {-1 / N_{\mathrm{c}}}\end{array}\right] \nonumber .\\
&&
\end{eqnarray}{}
\end{widetext}{}
The function $D(\vec{r})$ is related to the imaginary part of gluonic
self-energy. It reflects the scattering rate of off-shell ($\omega\ll|\vec{k}|$)
longitudinal gluons. For $r \sim 1/m_D$, the most important contributions
are captured by the Hard-thermal-loop(HTL) approximations
\cite{Kapusta:2006pm}. In this approximation,
\begin{equation}
\label{eqn:akadrHTL}
     D(\vec{r}) = -g^2 T\int \frac{d^{3} k}{(2 \pi)^{3}} \frac{\pi
     m_{\mathrm{D}}^{2} e^{i \vec{k} \cdot
     \vec{r}}}{k\left(k^{2}+m_{\mathrm{D}}^{2}\right)^{2}}\;.
\end{equation}{} 
Here $m_D$ is the Debye mass for which we use the one-loop result $m_D =
\sqrt{N_\mathrm{c}/3 + N_f/6}\ gT$. $N_f$ here is the number of light flavors. It is easy
to see that $D(\vec{r})$ approaches $0$ as $r$ increases beyond $1/m_D$. 

This master equation satisfies the necessary physical constraints of linearity,
positivity and trace-preservation. Techniques of quantum-state diffusion
methods \cite{gisin19921} can then be applied to numerically simulate the
evolution of such a master equation. For example,
Eq.~\ref{eqn:aka_reduced_density} can be simulated using the stochastic
evolution in the following manner \cite{akamatsu20151}.

One starts from the pure state at the initial time $t_0$ (although mixed states
can easily be used~\cite{brambilla20181}). One introduces noise
fields $\theta^a(t, \vec{r})$ which are picked from an ensemble which is
specified by the expectation values,

\begin{eqnarray}
\label{eqn:noise_full_correlation}
\langle\langle \theta^a(t, \vec{r}) \rangle\rangle &=& 0 \nonumber\\
\langle\langle\theta^a(t, \vec{r})\theta^b(t', \vec{r}') \rangle\rangle &=&
\delta^{ab}D(\vec{r}-\vec{r}')\delta(t-t')\;, 
\end{eqnarray}
where $\langle\langle .. \rangle\rangle$ means taking the stochastic average over the noise fields.

For each member of the ensemble $\theta^a(t, \vec{r})$, $\psi$ is evolved using
the Schr\"{o}dinger equation,
\begin{eqnarray}
\label{eqn:stochasticevolution}
\psi(t+dt) && = e^{-i H_{\theta}(t)dt} \psi(t)  \nonumber\\
H_{\theta}(\vec{r},t)=&&-\frac{\vec{\nabla}_{r}^{2}}{M} 
+ V(r)(t^{a}\otimes t^{a\ast}) \nonumber \\
&&+\theta^{a}(t,
\frac{\vec{r}}{2})\left(t^{a} \otimes 1\right)
-\theta^{a}(t,-\frac{\vec{r}}{2})\left(1
\otimes t^{a *}\right),\;
\end{eqnarray}
and the density matrix can be obtained by taking a stochastic average of the outer
product 
 \begin{equation}
 \label{eqn:stochastic_density_matrix}
 \rho(t, \vec{r}, \vec{s}) = \langle\langle\;\; 
    \ket{\psi(t, \vec{r}) }\bra{\psi(t, \vec{s})} \;\;\rangle\rangle\;.
\end{equation}

Master equations of similar form have also been solved in \cite{brambilla20171,brambilla20181,yao20181,yao20183} with different implementations.
A simplified version of Eq.~\ref{eqn:aka_reduced_density} was simulated in
\cite{akamatsu20181}, where the system was assumed to be one-dimensional and
the color-structure of $\qqb$ pair was neglected (Abelian dynamics). 

To incorporate these effects, we first simplify the stochastic evolution
equation (Eq.~\ref{eqn:aka_reduced_density}) (and therefore its corresponding
master equation) by expanding the decoherence terms in small $\vec{r},
\vec{s}$. This approximation is motivated by the hierarchy between the inverse
size of the states and the temperature $1/r\gg T$. 

This allows us to extend the calculation to a three dimensional system while
keeping all the color structure of $\qqb$ pair intact without a high
computational cost. The calculation is three dimensional in the sense that we
allow for transitions between different angular-momentum states ($l = 0,1$).
Transitions which change the angular-momentum by two units or more are
suppressed by $\mathcal{O}(r^2T^2)$. (See Ref.~\cite{brambilla20171} for a
similar analysis.)

To check the accuracy of $\vec{r}$ approximations, we performed a similar
expansion for Abelian dynamics in one dimension, for which results are known
from Ref.~\cite{akamatsu20181}. The comparison is presented in
Appendix~\ref{appex:appendix1}.  Without expanding in $\vec{r}$ we were able to
match the results of Ref.~\cite{akamatsu20181}, thereby testing our implementation. Then we
analyze conditions on the wavefunctions for which the $\vec{r}$ expansion is
accurate. The main conclusion from the analysis is that this is a good
approximation for the lowest two bound states of Bottomonia, and we focus on
these states in the three dimensional calculation.

\subsection{\label{subsec:decoherence13} Small $\vec{r}$ expansion and the momentum diffusion coefficient $\kappa$}
 
We start with the stochastic evolution equation for a quark-antiquark $\qqb$
pair in its rest frame (\ref{eqn:stochasticevolution}). The decoherence terms, in the
density matrix for $\qqb$ pair, are all expanded in small $\vec{r}$ expansion.

This simplifies the calculation in two ways. First, the noise field
$\theta^a(\vec{r}, t)$ correlated in space, is replaced by just two different
noises which are $\vec{r}$ independent and only depend on time. We only need
the noise field at the center-of-mass coordinate $\vec{R}$, and its first
derivative at $\vec{R}$. This makes generation of the stochastic noise much
cheaper computationally. Second, the $\vec{r}$ expansion allows one to compute
transitions between different angular momentum states, thus facilitating a
three-dimensional calculation.

The $\vec{r}$ expanded stochastic evolution operator up to
$\mathcal{O}(\vec{r}^2)$ for a $l=0$ initial state is,
\begin{eqnarray}
\label{eqn:paper_decoherence_1}
  && \psi(t+dt)  = e^{-i H_{\theta}dt}\psi(t) \nonumber \\  
  &&  H_{\theta} =  (\frac{-\nabla^2}{ M} (1_Q \otimes 1_{\bar{Q}}) + V(r)(t^a \otimes t^{\ast,a})  \nonumber \\
&& \qquad+D^{a} \ \frac{\vec{r}}{2}\cdot \vec{\theta}^{a}(t)+ F^{a} \  \theta^a(t)+\mathcal{O}(\vec{r}^2) ) \nonumber \\
&&\theta^{a}(t) = \theta^{a}(\vec{r},t)|_{\vec{r}=0},\quad \vec{\theta}_{i}(t)
= \vec{\nabla}_{i}\theta(\vec{r},t)|_{\vec{r}=0}\;,
\end{eqnarray}
where the noise field $\theta(\vec{r},t)$ was defined in
Eq.~\ref{eqn:noise_full_correlation}. ($F^{a}= (t^a_Q\otimes 1_{\bar{Q}} -
1_{Q}\otimes t^{\ast\;a}_{\bar{Q}})$ and $D^{a} = (t^a_Q\otimes 1_{\bar{Q}} +
1_{Q}\otimes t^{\ast\;a}_{\bar{Q}})$ are operators in the color-space of $\qqb$
pair. The subscript $i$ refers to the spatial index, and we refer
$\vec{\nabla}\theta$ at $\vec{r}=0$ as $\vec{\theta}$ for notational 
convenience.)

The noises appearing in Eq.~\ref{eqn:paper_decoherence_1} can be
generated as random-fluctuations correlated locally in time as,
\begin{eqnarray}
\label{eqn:paper_noise_correlation}
 && \langle\langle \theta^a(t)\theta^b(t')\rangle\rangle = \delta^{ab}\delta(t-t')D(\vec{0}),\nonumber \\
&& \langle\langle \vec{\theta_i}^a(t) \vec{\theta_j}^b(t')\rangle\rangle = \delta^{ab}\delta(t-t')\delta_{ij}\frac{-\nabla^2}{3}D(\vec{0}).
\end{eqnarray}{}
Note that the factors of $g$ have been absorbed in the definition of the
correlation function (Eq.~\ref{eqn:noise_full_correlation}). The above Hamiltonian evolution is written for a three
dimensional system. Since, $V(\vec{r})$ is rotationally invariant, we can separate
the radial part of the three dimensional wavefunction from its angular part. The wavefunction
in position space can be written as

\begin{equation}
    \label{eqn:3d_wavefunctions}
    \Psi(\vec{r},t) = \frac{\psi(r)}{r}\upvartheta(\beta,\phi),
\end{equation}
where $\psi(r)$ is the radial wavefunction and $\upvartheta$ is the wavefunction in angular momentum space, with $\beta$ being the polar angle and $\phi$ azimuthal angle. We also define the normalized color states for $\qqb$ octet and singlet wavefunction as,
\begin{equation}
\label{eqn:color_wavefunctions}
\ket{S} = \frac{\mathds{1}}{\sqrt{N_c}}\sum_{lk}\ket{lk} \quad \ket{O^a} =  \frac{1}{\sqrt{T_F}}\sum_{lk} (t^a)_{lk}\ket{lk}.
\end{equation}
The indices $l,k$ denotes the color states of a single quark or antiquark. $T_F = 1/2$ is the index for $SU(3)$ Lie group, for the fundamental representation. 

Finally, we project the evolution operator in the Eq.~\ref{eqn:paper_decoherence_1} into the color and angular momentum space of $\qqb$ pair,
\begin{figure}
\includegraphics[height=6cm,width=\linewidth]{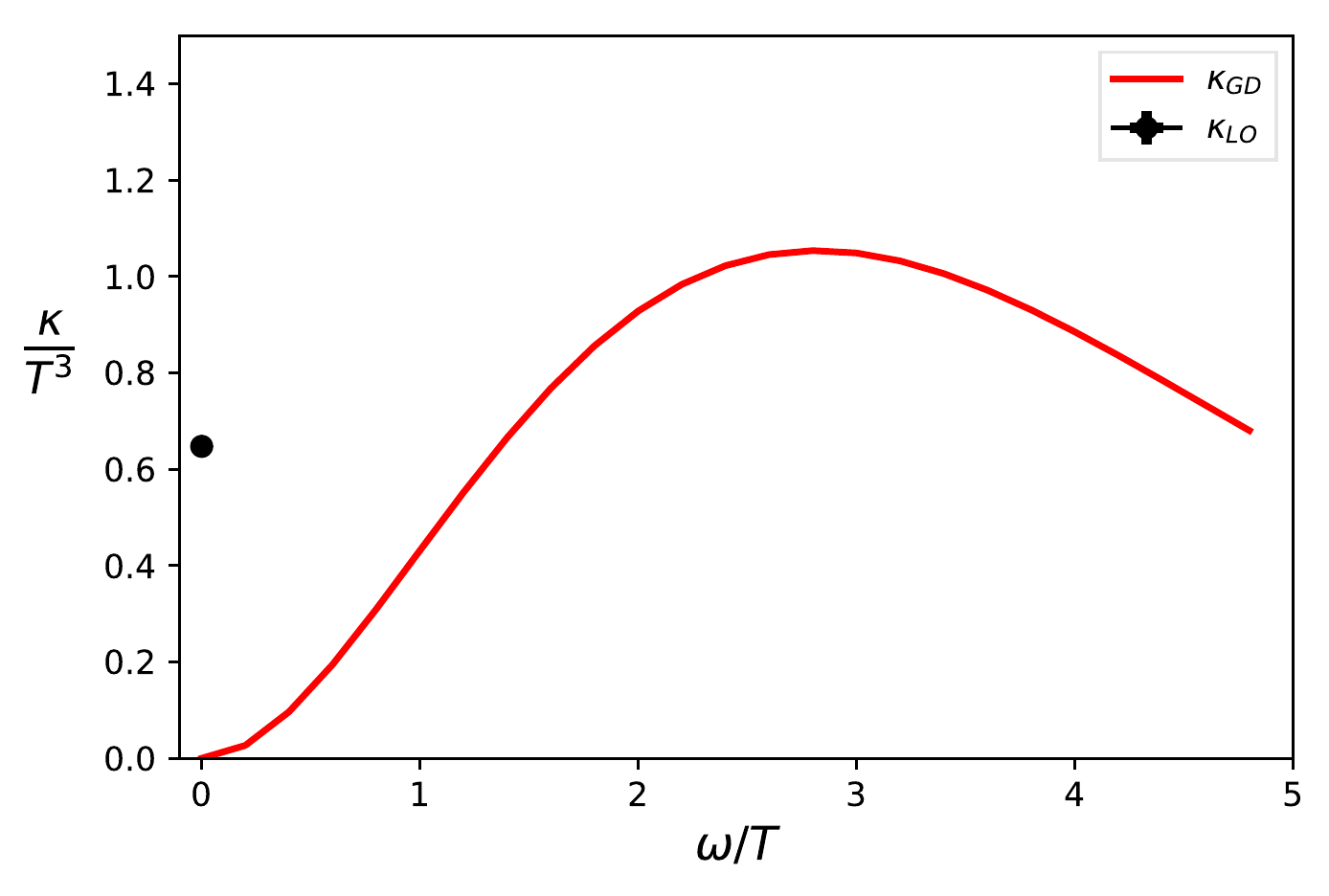}
\caption{\label{fig:kappa_comparison} (color online) Comparison of $\kappa$ between its LO
value \cite{huot20081} and the correlator $\Tilde{G}(\omega,T)$(see Eq. 
\ref{eqn:EE_finite_frequency}), since it plays a similar role for the
calculation done in Sec.~\ref{sec:pNRQCD}. See the value of
$\kappa_{\mathrm{Lattice}}$ also in text. The value of $g = 2.27$ was used.}
\end{figure}
\begin{widetext}
\begin{equation}
\label{eqn:paper_matrix_eqn1}
H_{\theta}(r,t) =  \left(\begin{array}{cccc}
H^{S}_{0}(r,t) & 0 & 0 & \frac{1}{\sqrt{2N_c}} r|\vec{\theta}^{c}(t)|\delta_{ac} \\
0 & H^{S}_{1}(r,t) & \frac{1}{\sqrt{2N_c}} r |\vec{\theta}^{c}(t)|\delta_{ac} & 0 \\
0 & \frac{1}{\sqrt{2N_c}} r |\vec{\theta}^{c}(t)|\delta_{ac} & H^{O}_{0}(r,t)+f^{abc}\theta^{c}(t)& \frac{d^{abc}}{2} r |\vec{\theta}(t)| \\
\frac{1}{\sqrt{2N_c}} r|\vec{\theta}^{a}(t)|\delta_{ac} & 0  &\frac{d^{abc}}{2}r|\vec{\theta}(t)|&  H^{O}_{1}(r,t)+f^{abc}\theta^{c}(t)  
\end{array}{} \right). 
\end{equation}
\end{widetext}

This Hamiltonian acts on the wavefunction given in the form 
 \begin{equation}
 \label{eqn:wavefunctions_section1}
 \psi(r,t)  = \left(\psi^{S}_{l=0}(r,t), \psi^{S}_{l=1}(r,t),\psi^{O^{a}}_{l=0}(r,t),
      \psi^{O^{a}}_{l=1}(r,t) \right).
 \end{equation}
 
Here, $\psi^S(r,t)$ and $\psi^{O^a}(r,t)$ denote radial wavefunctions
for $\qqb$ pair in singlet and octet states respectively and the index $a$ runs
from $1$ to $(N_c^2-1)$ for different color-octet states. $l$ denotes the angular
momentum states, which take the values $l = 0,1$. The Hamiltonians for the
singlet and octet states are 
\begin{eqnarray}
\label{eqn:section1_hamiltonians}
H^S_l && = -\frac{1}{M}\frac{\partial^2}{\partial r^2} - \frac{C_{\mathrm{F} }\alpha}{r}e^{-m_D r} + \frac{l(l+1)}{ M r^2} \nonumber, \\
H^O_l && = -\frac{1}{M}\frac{\partial^2}{\partial r^2} + \frac{\alpha}{2 N_{\mathrm{c}}r}e^{-m_D r} + \frac{l(l+1)}{M r^2}. 
\end{eqnarray}{}
One can check that the color factors between singlet and octet states are same as those obtained in pNRQCD \cite{pineda19981,brambilla20001}.

Under the approximations considered, the correlation functions
(Eqs.~\ref{eqn:paper_noise_correlation}) are the most important quantities
which control the suppression pattern.  In Ref.~\cite{akamatsu20181}, the
correlation function (Eq.~\ref{eqn:noise_full_correlation}) was approximated by a
gaussian function with a width $l_{\mathrm{corr}}\sim gT$.  Here we simply use
the HTL form (Eq.~\ref{eqn:akadrHTL}). (In Appendix~\ref{appex:appendix1} we use the
gaussian form since we wanted to compare with Ref.~\cite{akamatsu20181}.)
$-\vec{\nabla}^2 D(0)$ at one loop HTL is
divergent. This problem is well known in the perturbative calculations of
momentum-diffusion coefficients for a heavy-quark \cite{braaten19911}. This is
not physical as the problem arises from the use of the HTL form for
$D(\vec{r})$ for very short distances where it is not valid.  For the momentum
diffusion coefficient, 
\begin{equation}
\label{eq:kappa_vs_D0}
\kappa = -\frac{C_F}{3} \nabla^2 D(\vec{r})|_{\vec{r}=0}
\end{equation}
the contribution from the scales above $gT$ is important. The resulting
ultraviolet divergence in the soft-momentum region which is regulated by a
cutoff of the order of $gT$ is cancelled by the infrared divergence coming from
the upper momentum sector $k\sim gT-T$ (see \cite{braaten19911, moore20081,
brambilla20081}). Since the constant $\kappa$ is closely related to the
physical observable such as the flow patterns of heavy quarks inside the QGP
medium, it has been investigated extensively. In our calculation, we use the
weak coupling result for $\kappa$ at the leading-order (LO) including the UV
contributions~\cite{solana20061,huot20081}, given in
Eq.~\ref{eqn:kappa_LO_huot},
\begin{eqnarray}
    \label{eqn:kappa_LO_huot}
    && \kappa^{L O} = \frac{g^{4} C_{F}}{12 \pi^{3}} \int_{0}^{\infty} k^{2} d k \int_{0}^{2 k} \frac{q^{3} d q}{\left(q^{2}+m_{D}^{2}\right)^{2}} \nonumber \\
    && \qquad \quad \times\left\{\begin{array}{l}{N_{\mathrm{c}} n_{B}(k)\left(1+n_{B}(k)\right)\left(2-\frac{q^{2}}{k^{2}}+\frac{q^{4}}{4 k^{2}}\right)} \\ {+N_{f} n_{F}(k)\left(1-n_{F}(k)\right)\left(2-\frac{q^{2}}{2 k^{2}}\right)}\end{array}.\right.
\end{eqnarray}
The value of $\kappa$ has also been calculated on
lattice \cite{datta20121,laine20151} for a pure $SU(3)$ theory and it was seen
to be larger compared to its LO estimates
from perturbative calculations,$ 1.8 \lesssim \kappa_{\mathrm{Lattice}}/T^3 \lesssim
3.4$. Including light quarks in the calculation might modify this value
further. 

Intuitive understanding of the noise field can be gleaned by looking at the
non-perturbative expression for the diffusion constant in terms of the
correlation function of color-electric fields  at different times
\cite{solana20061},
\begin{eqnarray}
\label{eqn:momentum_diffusion_solana}
    && \kappa = \frac{g^{2}}{3 N_C} \int_{-\infty}^{\infty} 
    d t \tr_{H}\left\langle W(t ;-\infty)^{\dagger} 
    E_{i}^{a}(t) t^{a} W(t ; 0) \nonumber \right. \\
    && \left.\qquad \qquad \times E_{i}^{b}(0) t^{b} 
    W(0 ;-\infty)\right\rangle,
\end{eqnarray}
where $E_{i}^{a}(t)$ is the color electric field and $W(t ; 0)$ is the
gauge link in the fundamental representation. 

The small $\vec{r}$ expansion of $D(\vec{r})$ gives us exactly the same
quantity which one uses in these sorts of perturbative calculation (see
\cite{solana20061,huot20081,datta20121, akamatsu20151}).

Comparing Eqs.~\ref{eqn:momentum_diffusion_solana},~\ref{eq:kappa_vs_D0} with  with
Eq.~\ref{eqn:paper_noise_correlation} we see that $\nabla\theta^a$ can
simply be interpreted as $\vec{E}^a$ in the temporal gauge except for the
factors of $g$ which has been absorbed in the definition of noise correlations. 

It was shown in Ref.~\cite{brambilla20081} that the same structure of the
electric-field correlator appears in the calculation of the gluo-dissociation
rate, where the gauge link connecting the two fields is in the adjoint space.
In temporal gauge the two different correlator defined perturbatively becomes
same. Therefore we have also plotted the relevant correlator (see Eq.
\ref{eqn:electricfield_in_decayrate} and denoted as $\kappa_{GD}$ on the plot)
on the same plot (Fig. \ref{fig:kappa_comparison}). Therefore one can argue
that noise-fields here can be thought of as the electric field present in the
pNRQCD lagrangian \cite{pineda19981,brambilla20001}. We can extend the
definition of the $\vec{\theta}^a(t)$ correlator from being uncorrelated in
time to have a finite correlation in time to include on-shell processes. This
modification ensures the gluonic emission and absorption processes are include
in our calculation. This we do next.

\section{\label{sec:pNRQCD}Gluo-dissociation}
In this section, we describe our implementation of the quantum calculation of
the process called
gluo-dissociation~\cite{peskin19791,bhanot19791,xu19951,brambilla20081,brambilla20111} in
literature. At finite temperature, a singlet bound state can absorb a gluon
from the medium and jump to one of the excited state. This process changes the
color state of the quarkonia to a color-octet state. In perturbation theory,
the short distance potential for an octet state is repulsive and thus it is
typically assumed that this transition destroys the bound state. 

The decay rate from this process assuming $T\ll 1/r$ was first calculated in
Refs.~\cite{peskin19791,bhanot19791}. More recently, the decay rate was
computed for $T\gg E_b$ in Ref.~\cite{brambilla20081} and $1/N_c$ corrections for
$T\ll 1/r$ were computed in Ref.~\cite{brambilla20111}.

The process of gluo-dissociation is naturally described in
pNRQCD~\cite{pineda19981,brambilla20001}. This EFT is valid at energies (this
could be an energy scale like $T$, $m_D$ or $E_b$ depending on the hierarchies
between these scales) $\ll 1/r$. The degrees of freedom in the theory are light
degrees of freedoms like gluons and light quarks, and the singlet and octet
wavefunctions of the $\qqb$. (For a detailed study of pNRQCD at finite $T$ see
\cite{brambilla20081}.)

Starting with the lagrangian,

\begin{eqnarray}
\label{eqn:pNRQCDlag}
&&\mathcal{L}_{\mathrm{pNRQCD}}=
-\frac{1}{4} F_{\mu \nu}^{a} F^{a \mu \nu}
+ \sum_{i=1}^{N_{f}} \overline{q}_{i} i \slashed{D} q_{i} 
\nonumber \\
&& \quad+ \int d^{3} r \ 
\tr \left\{\Psi^{\dagger,S}
\left[i \partial_{0}-h_{s} \right] \Psi^{S}
+\Psi^{\dagger,O}\left[i D_{0}-h_{o}\right]\Psi^{O}\right. 
\nonumber \\
&&\quad + V_{A}\left(\Psi^{\dagger,O}\vec{r} 
\cdot g \vec{E}\Psi^{S}+\mathrm{H.c.}\right)
+\frac{V_{B}}{2} \Psi^{\dagger,O}\{\vec{r}
\cdot g \vec{E}, \Psi^{\dagger,O}\}
\nonumber \\
&&\quad +\left. \ldots \right\}  ,
\end{eqnarray}
where $\Psi^{S}(\vec{r},t) = \Psi(\vec{r},t)\otimes \ket{S}$ and $\Psi^{O}(\vec{r},t)
= \Psi^{a}(\vec{r},t)\otimes \ket{O^a}$ are the three dimensional wavefunctions
of $\qqb$ pair in color-singlet and octet states, where color states $\ket{S}$
and $\ket{O^a}$ has been defined in Eq.~\ref{eqn:color_wavefunctions}.
$\partial_{0}$ denotes the time derivative and $D_{0} = \partial_{0} - i g
A_{0}^{a}t^{a}$ is the covariant derivative acting on the octet states. The
Hamiltonian for singlet and octet states are  
\begin{eqnarray}
\label{eqn:section3_hamiltonians}
h^S && = -\frac{\vec{\nabla}^2}{M} - \frac{C_{\mathrm{F}}\alpha}{r}e^{-m_D r}  ,\nonumber \\
h^O && = -\frac{\vec{\nabla}^2}{M} + \frac{\alpha}{2 N_{\mathrm{c}}r}e^{-m_D r}.
\end{eqnarray}{}

The interaction vertices are same as in Eq.~\ref{eqn:paper_matrix_eqn1}.
$F_{\mu \nu}^{a}(\vec{R},t)$ is the field strength tensor for the long
wavelength gluons and $q(\vec{R},t)$ represents the light
quarks. Because of the multipole expansion of pNRQCD, all the light degrees of
freedom are function of center-of-mass coordinate ($\vec{R}$) only.  $V_{A}$
and $V_{B}$ are the coefficients of dipole-interactions. At leading order in
$\alpha$ they are $1$.

The wavefunctions $\Psi^{S}, \Psi^{O}$ in Eq.~\ref{eqn:pNRQCDlag} are
different from the wavefunctions $\psi^{S},\psi^{O}$ in Eq. 
\ref{eqn:wavefunctions_section1} as they are for a three dimensional system
right now. One can project out the above lagrangian in the color and angular
momentum space of $\qqb$ to get back to an equation similar to Eq. 
\ref{eqn:paper_matrix_eqn1}. 
The singlet to octet transition rate in a thermal medium at a uniform, time
independent, temperature $T$ to first order in perturbation theory is
\begin{eqnarray}
\label{eqn:decayratesing}
\Gamma_{\psi_S} = \int_{\psi_{\mathrm{O}}}
|\braket{\psi_{\mathrm{O}}|\vec{r}|\psi_{\mathrm{S}}}|^2 \Tilde{G}(\Delta E,T).
\end{eqnarray}{}
\begin{figure}[h]
\includegraphics[width = 0.5\textwidth]{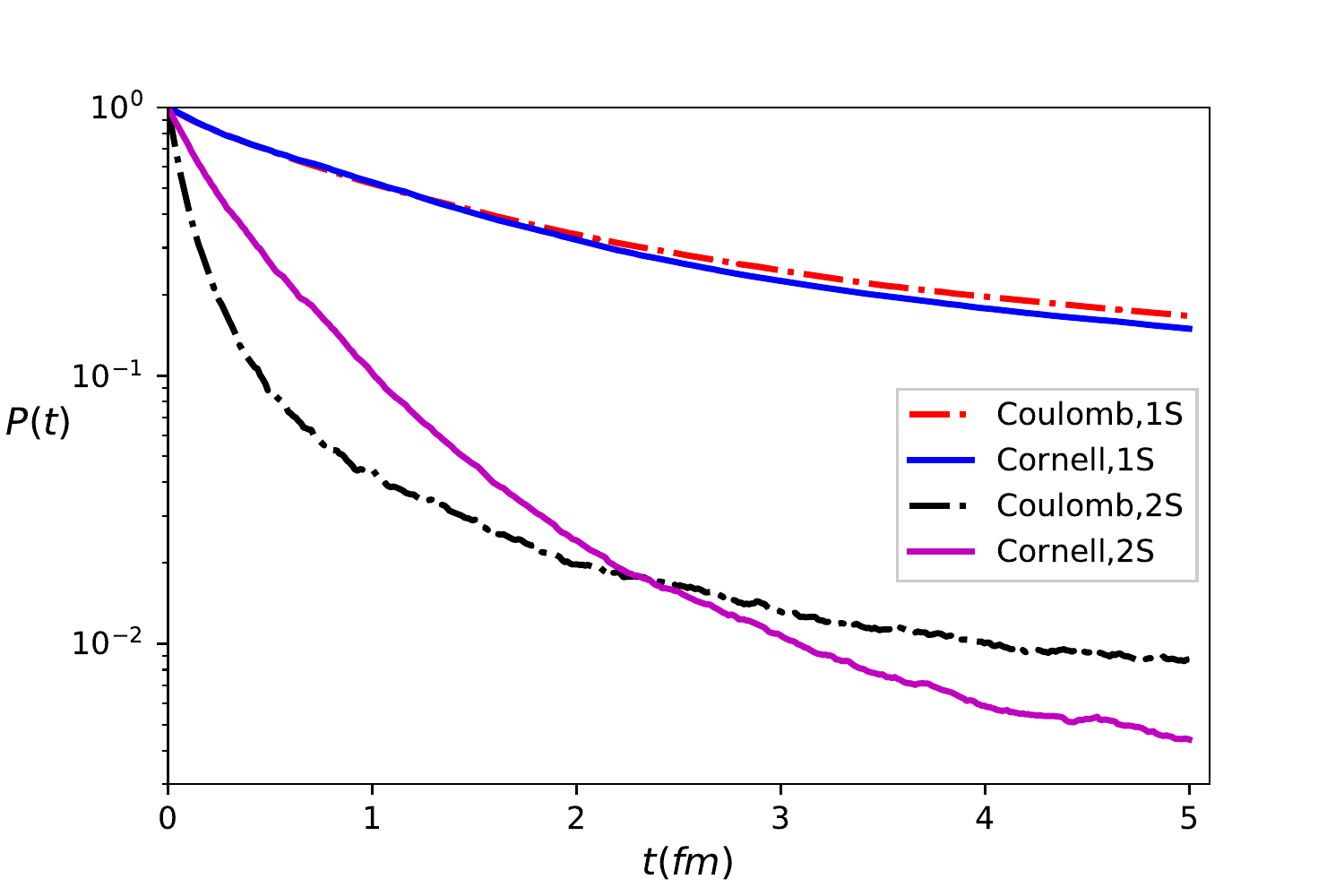}
\caption{(color online) Comparison of survival probability $P(t)$ for 
decoherence (Eq.~\ref{eqn:paper_noise_correlation}).  For $1S$ states (blue solid curves for $1S$ eigenstates of the
Cornell potential as initial states and red dot dashed for Coulomb) the
suppression is very similar except at final times. $2S$ states (pink solid for Cornell
and black dot dashed for Coulomb) show a more interesting behavior as they are
affected by the potential change very strongly.}
\label{fig:final_results_decoherence}
\end{figure}
The integration is over the set of continuum of octet states. We will focus on
$l=0$ singlet initial states and therefore the final octet states have $l=1$.
$\Delta E = E(\psi_O) - E(\psi_S)$. $\Tilde{G}(\omega,T)$ is given by
the thermal expectation value
\begin{eqnarray}
    \label{eqn:electricfield_in_decayrate}
    &&\Tilde{G}(\omega,T) = \nonumber \\
    && \quad \frac{g^2\pi}{3 N_{\mathrm{c}}}\tr\Bigl\{e^{-\mathcal{H}/T}
     \int_{-\infty}^{\infty} d{t}
     \vec{E}^{a}(\vec{R},t)\phi^{ab}(t,0)\vec{E}^{b}(\vec{R},0)
      e^{i \omega t}\Bigr\} \nonumber \\&&
\end{eqnarray}
Where $\phi^{ab}(t)$ is the gauge link connecting the two electric-fields in
adjoint representation~\cite{brambilla20081}.

Looking at Eq.~\ref{eqn:electricfield_in_decayrate}, we see that it has the
same structure as that of the correlator in
Eq.~\ref{eqn:momentum_diffusion_solana} with two important differences. First,
the gauge link is adjoint in Eq.~\ref{eqn:electricfield_in_decayrate} and
fundamental in Eq.~\ref{eqn:momentum_diffusion_solana}. Second,
Eq.~\ref{eqn:electricfield_in_decayrate} has an additional factor of
$e^{-i\omega t}$ corresponding to the fact that during the gluo-dissociation 
process, the $\qqb$ state absorbs energy $\omega$ from the gluon. 

The absence of $e^{-i\omega t}$ in Eq.~\ref{eqn:momentum_diffusion_solana} can
be traced to the hierarchy between the energy scales assumed in the derivation
of Eq.~\ref{eqn:aka_reduced_density}. $E_b\ll gT$ implies that the relaxation
time scale for the thermal gluons is much shorter than the system time scales.
Therefore, the electric field correlator can be taken to be local in
time on long time scales. Physically it corresponds to
the assumption that there are no long-lived (compared to $1/E_b$) gluonic degrees 
of freedom in the medium. 

\begin{figure}[h]
\includegraphics[width = 0.5\textwidth]{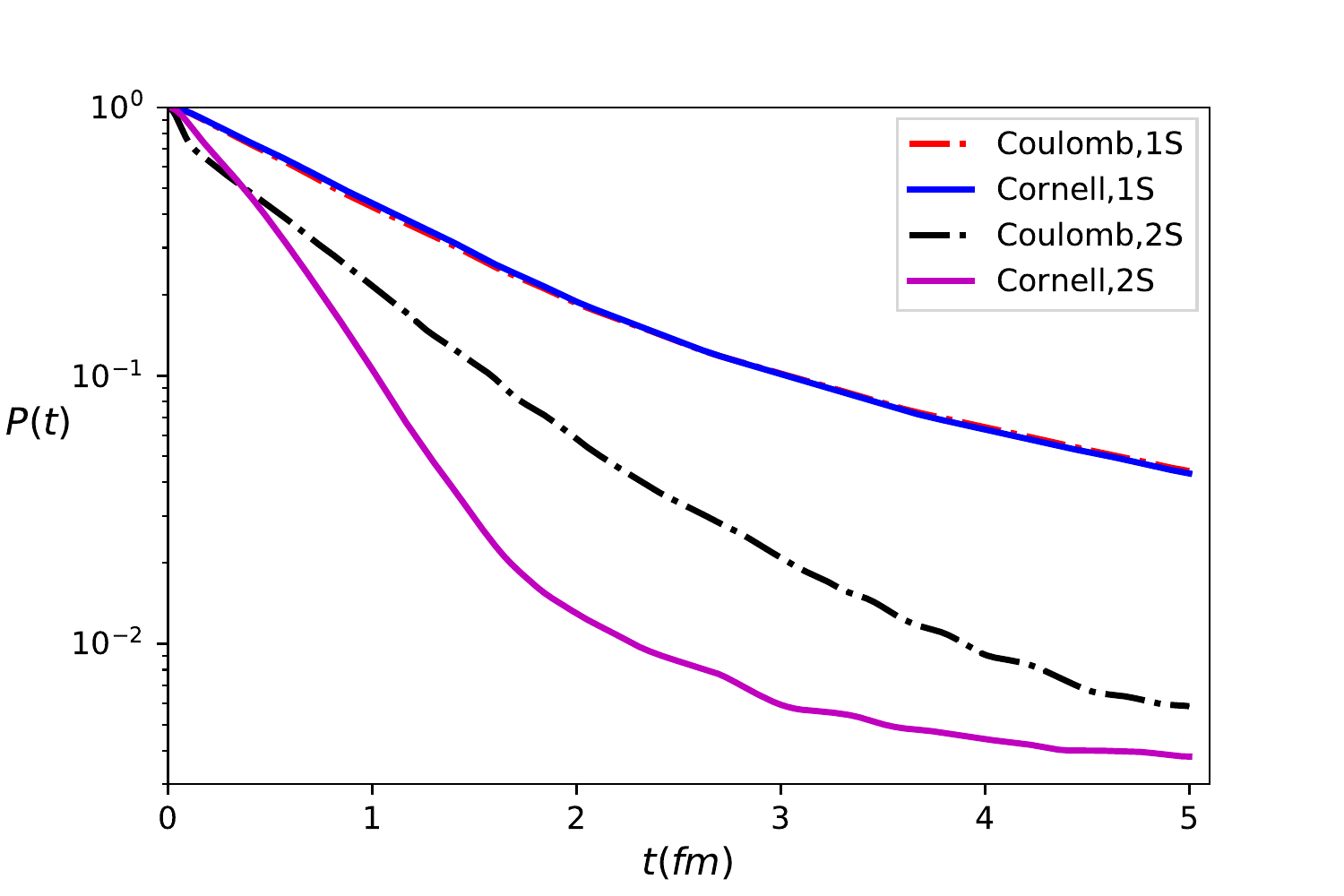}
\caption{\label{fig:final_results_gluodissociation}
(color online) Comparison of $P(t)$ for gluo-dissociation (Sec.~\ref{sec:pNRQCD}).
For $1S$ states (blue solid for Cornell and red dot dashed for Coulomb) the
suppression is very similar at all times. $2S$ states (pink solid for Cornell and black
dot dashed for Coulomb) show that the evolution of survival probability is non
trivial. Comparing this with Fig.~\ref{fig:final_results_decoherence}, we see that their behavior is different for two different medium effects.}
\end{figure}
On relaxing this assumption, $\omega$ can no longer
be taken to be zero in Eq.~\ref{eqn:electricfield_in_decayrate}, and the 
electric field correlator has a finite correlation in time. In the calculation 
of the decay rate this does not cause any technical complication
(Eq.~\ref{eqn:decayratesing}). However, in a quantum calculation which
follows the density matrix evolution of the $\qqb$, the steps involved in deriving a
Markovian evolution in the form Eq.~\ref{eqn:aka_reduced_density} can no longer
be followed. 

As an illustrative example, consider a regime when the relaxation rate of the
thermal gluons is small compared to $T$, and let $T$ be comparable to $E_b$.
Concretely, one scenario where this can be realized in the weak coupling regime
when the gluonic screening mass, $\sim gT$, and the relaxation rate, $\sim
g^2T$,~\cite{Kapusta:2006pm} are both much smaller than $T$. Then, at leading
order in $g$, the electric field correlator can be written as,  
\begin{eqnarray}
\label{eqn:EE_finite_frequency}
  &&\tr \langle e^{-\mathcal{H}/T}[g\vec{E}^a_i(t)][g\vec{E}^b_j(t')]\rangle \nonumber \\
  &&\qquad = 
  \delta_{ab}\delta_{ij}\frac{g^2 T^4}{6 N_{\mathrm{c}}\pi} 
  \int_{0}^{\infty} d\xi \ x^3 \cos(\xi\;T\;(t-t')))\frac{1}{e^{\xi}-1} 
  . \nonumber \\
 &&
\end{eqnarray}{}
where $a$, $b$ are color indices and $i$, $j$ are spatial indices.

In this case the thermal gluons can not be integrated out from the influence
functional to obtain an interaction term which is local in time.  To make
progress on the quantum implementation in presence of a correlated electric
field, we start from the stochastic Schr\"{o}dinger Eq. \ref{eqn:paper_matrix_eqn1}. Following the interpretation in 
Sec.~\ref{sec:decoherence1} of the noise field $\vec{\theta}^a(t)$ as
$g\vec{E}^a(t)$, the correlation function of stochastic noise is given as

\begin{eqnarray}
\label{eqn:noise_generation_eqlb_finite_frequency}
 \langle\langle \vec{\theta}^a(t) \rangle\rangle &&= 0 \nonumber\;. 
\end{eqnarray}{}
The correlator $\langle\langle  \vec{\theta_{i}}^a(t)\vec{\theta_{j}}^a(t) \rangle\rangle$ 
is given by Eq.~\ref{eqn:EE_finite_frequency}.

The density matrix at any given time can be obtained by taking the noise
average Eq.~\ref{eqn:stochastic_density_matrix}. The evolution equation for the
density matrix thus obtained can not be written in a Markovian form as the
correlations between the noise terms are not local in time.

A more rigorous approach to obtaining a time evolution equation for the density
matrix would involve deriving the influence functional without making an expansion in
$\omega$, and using the full gluonic propagator. Here we have used the lowest
order form for the electric-field correlator
(Eq.~\ref{eqn:EE_finite_frequency}).  At one-loop the spectral function of
gluons changes drastically (see fig.~\ref{fig:kappa_comparison})
for the particles with momenta less than $T$. The finite thermal mass and decay
width is important and can not be ignored. These corrections will change the
spectral density and also the analysis of non-Markovian regime. Finally
spontaneous emission processes need to be included. We leave these
considerations for future work. In spirit, our calculation is similar to what
was done in Ref.~\cite{akamatsu20121}, before it was made theoretically concrete in
subsequent works Refs.~\cite{akamatsu20131,akamatsu20151}. 

At this point we would like to make a comment about an alternative approach to
deriving the quantum evolution equations for quarkonia in the QGP. Open quantum
treatment of quarkonia starting from the pNRQCD lagrangian has been performed in
\cite{brambilla20171,brambilla20181}. The authors derived a general evolution
equation for the $\qqb$ density matrix including gluo-dissociation processes.
Furthermore, in two different physical regimes, they were able to write the
density matrix equations in Lindblad from. The first case was the strong 
coupling regime, in which case the static limit of the electric field 
correlator was considered. The second case was the weak coupling limit,
$g\ll1$, where the leading order form for the electric-field correlator was
used just as in Eq.~\ref{eqn:EE_finite_frequency}. The evolution equations were
simplified using the hierarchy $E_b\ll T$. In this case an expansion in
$(V_o-V_s)/T$ is possible, and the self-energy correction and the
gluo-dissociation rate can be simplified. We do not make this assumption. As a
result we can not write a simple equation for the density matrix evolution and
prove its validity in the weak coupling regime. 

However, we believe that this is a good first step towards incorporating
non-Markovian effects in the $\qqb$ evolution equations in the presence of
long-lived gluonic degrees of freedom.  This can also be confirmed by looking
at the classical decay picture with the quantum one at early time. We confirm in
Sec.~\ref{sec:final_results} that for small time, both results follow each
other and the two approach start diverging at late times (the details depend on
the initial states chosen, see \ref{sec:final_results}).

In a medium evolving with time we can modify the generation of noise to
incorporate the dependence of $T$ on time. Assuming that temperature change is
slow enough, we can approximate the physical picture as follows.  
\begin{enumerate}
    \item{The entire evolution of the $\qqb$ pair is divided into time blocks.
    During each block we take the $T$ to be constant and equal to the mean
    temperature in the block.  The division has to be done while keeping in
    mind that the time blocks we choose are large enough to include the finite
    correlation for the dominant gluons (which here are of the order $\sim
    T$).}
    \item{Suppose the time-interval is divided into $\beta$ number of blocks -
    $(0,t_{f}) = ((0,t_{1}),(t_{1},t_{2})\ldots,(t_{f-1},t_{f}))$, at
    $\beta$'th block the temperature is chosen to be $T_{\mathrm{eff}}(\beta) =
    (T(t_{\beta+1})+T(t_{\beta}))/2$. Where $T(t)$ is calculated by assuming a
    Bjorken evolution of the medium (see Eq.~\ref{eqn:temp_bjorken_evolution}) .}
    \item{Then for each block, the noise is generated using the equilibrium
    correlation function. They are stitched together using a linear
    interpolation functions, $\lambda_{\beta}(t)$ in Eq.
    ~\ref{eqn:noise_generation_noneqlb_finite_frequency}, which are normalized
    to one.} 
\end{enumerate}

We have used $3$ time-blocks for the results presented in
Sec.~\ref{sec:final_results}. We have also checked that using $5$ blocks
gives the same results. The stitching was done as given below.

\begin{eqnarray}
\label{eqn:noise_generation_noneqlb_finite_frequency}
&& \langle\langle \vec{\theta}^a_{i,\beta}(t)\rangle\rangle = 0 ,\nonumber \\ 
&&\langle\langle \vec{\theta}^{a}_{i,\alpha}(t)\vec{\theta}^b_{j,\beta}(t')\rangle\rangle 
 = \delta_{ij}\delta_{ab}\delta_{\alpha\beta} G(t-t', T_{\beta}) ,\nonumber \\
 &&\vec{\theta^{a}}(t) = \sum_{\beta}\lambda_{\beta}(t)\vec{\theta}^{a}_{\beta}(t),\quad \sum_{\beta}\lambda_{\beta}(t) = 1 .
\end{eqnarray}{}
Here, $\vec{\theta}_{i,\beta}^{a}(t)$ denotes the $i$'th component of noise 
$\vec{\theta}^{a}(t)$ generated in $\beta$'th block and $G(t,T)$ is given by 
Eq.~\ref{eqn:EE_finite_frequency}.

Having described these two different decay mechanisms and our implementation,
we proceed to the next section where we present our main results.

\section{\label{sec:final_results} Results}
The main results of the paper are presented here. We calculate the survival
probability of the vacuum states by doing a three dimensional quantum
evolution of the density matrix using the stochastic Schr\"{o}dinger equation
defined in the Eq.~\ref{eqn:paper_matrix_eqn1}, for two different physical
cases (decoherence and gluo-dissociation). The survival probability is defined
as 
\begin{equation}
    \label{eqn:survival_probability}
    P(t) = \langle\langle \  |\braket{\psi_{0}|\psi_{\theta}(t)}|^2  \ \rangle\rangle,
\end{equation}
where $\ket{\psi_{0}}$ is the vacuum wavefunction for $1S$ or $2S$ states.
The evolved wavefunction is 
\begin{equation}
    \label{eqn:final_section_wavefunction_time_evolution}
    \ket{\psi_{\theta}(t)} = e^{-i\int dt
H_{\theta}(t)}\ket{\psi_{0}}.
\end{equation}
The quantity $P(t)$ is related to the observed
suppression number $R_{AA}$ of quarkonium states at RHIC and LHC.
\begin{figure}[h] 
\includegraphics[width = 0.5\textwidth]{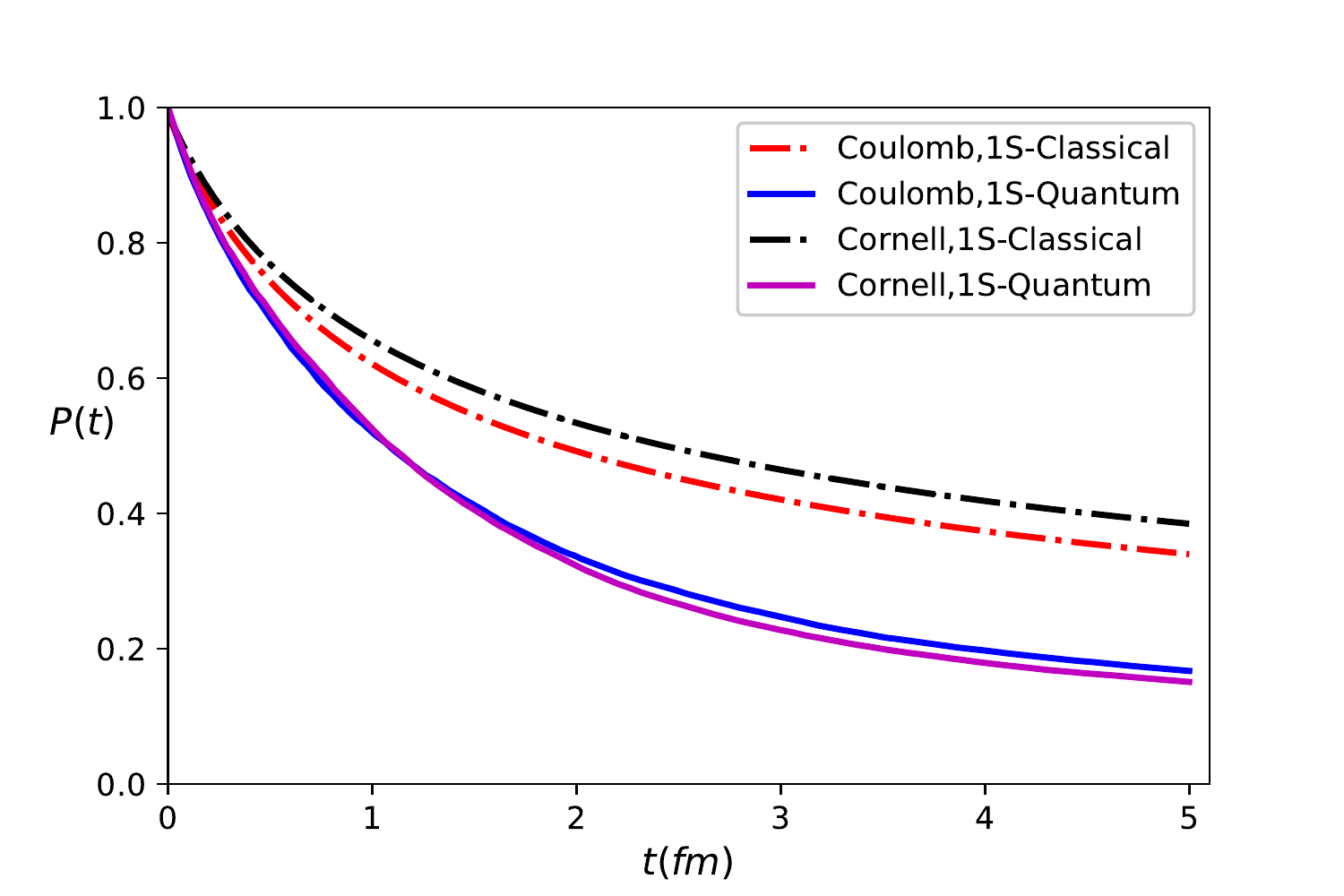}
\caption{\label{fig:final_results_qvsc_decoherence}
(color online) Comparison of $P(t)$ between
classical (black dot dashed for Cornell and red dot dashed for Coulomb) and
quantum (pink solid for Cornell and blue solid for Coulomb) approach for the
case of decoherence. The classical results start to differ as early as $t
\simeq 1\mathrm{fm}$ for the two initial wavefunctions whereas the quantum
results follow each other.}
\end{figure}
Typically, phenomenological calculations of $R_{AA}$ in the literature (see
\cite{grandchamp20041,krouppa20151,krouppa20171,krouppa20181}
use a classical rate-equation approach. We call these approaches
``classical'', since a full quantum evolution of the density matrix is not done.
Quarkonia has a finite decay width inside a thermal medium due to different
physical processes, such as inelastic scattering with thermal particles,
gluo-dissociation etc. The width can be calculated in perturbation theory at
desired order. Suppose there were initially $N_{\psi}$ numbers of
quarkonia in some state, labelled as $\psi$ here. The number of
surviving quarkonia after a finite time $t$ in the state $\psi$, in the classical
approach is given by 
\begin{eqnarray}
    \label{eqn:classical_RAA}
    && \frac{d N_{\psi}(t)}{dt} = - \Gamma_{\psi}(t) N_{\psi}(t) ,\nonumber \\
    && N_{\psi}(t) = e^{-\int_{t_{0}}^{t} dt' \Gamma_{\psi}(t')}N_{\psi}(t_{0}).
\end{eqnarray}

The value of $\Gamma_{\psi}(t)$ also depends on the choice of wavefunction one
uses to calculate the width. For example in
\cite{krouppa20151,krouppa20171,krouppa20181}, instantaneous value of the width
was used to calculate $N_{\psi}(t)$ by solving the three-dimensional
Schr\"{o}dinger equation at each time step using a complex potential. Its time
dependence comes from the fact that dissociation rate depends on $T$. The
quantity $P(t)$ defined in Eq. \ref{eqn:survival_probability} is equivalent to
$N_{\psi}(t)$ defined above in the sense that starting from $N$ $\qqb_{\psi}$
pair in state $\psi$ at time $t_{0}$, the number of surviving pair after time
$t$ is $P(t)\times N$. Therefore, from here on we use $P(t)$ to denote both the
quantum and classical survival probability.

In our calculation, we use the vacuum wavefunction at each time step to
calculate the value of the width. The expression for decay width for two
different cases (subscript '$dc$' for decoherence and '$gd$' for
gluo-dissociation) is 

\begin{eqnarray}
    \label{eqn:classical_width_dc}
    && \Gamma_{dc}(\psi, t) = \int_{\psi_{f}} |\bra{\psi_{f}}\hat{r}\ket{\psi_{0}}|^2\times \kappa(t) ,\nonumber \\
    &&\Gamma_{gd}(\psi, t) = \int_{\psi_{f}} |\bra{\psi_{f}}\hat{r}\ket{\psi_{0}}|^2\times \Tilde{G}(E_{f}-E_{i},t),
\end{eqnarray}
\begin{figure}[h]
\includegraphics[width = 0.5\textwidth]{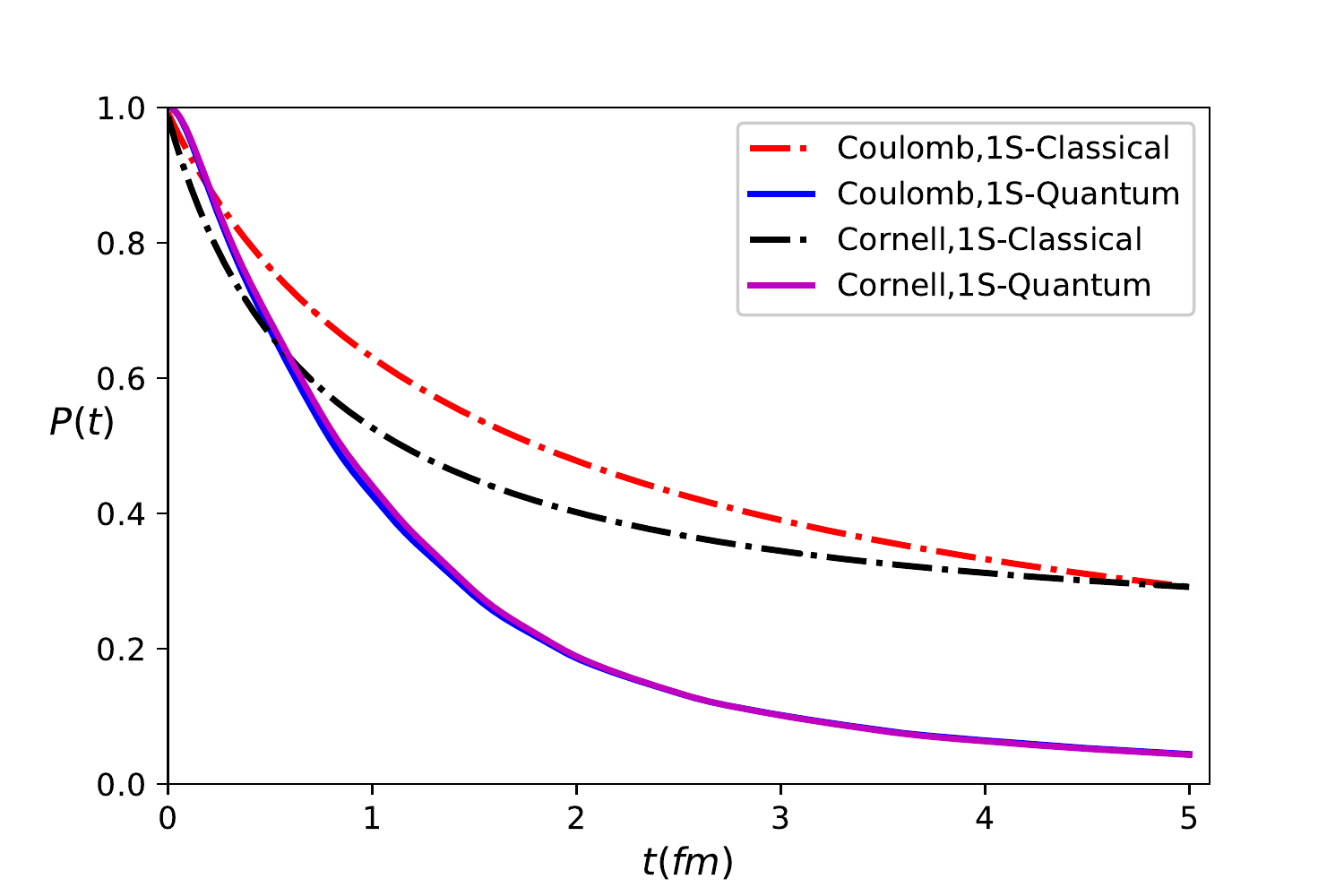}
\caption{\label{fig:final_results_qvsc_gluodissociation}
(color online) Comparison of $P(t)$ between
the classical (black dot dashed for Cornell and red dot dashed for Coulomb) and
quantum (pink solid for Cornell and blue solid for Coulomb) approach for the
case of gluo-dissociation. We see the same behaviour as in
Fig.~\ref{fig:final_results_qvsc_decoherence}. The classical results at early
time differ substantially for the two initial wavefunctions we used, whereas
the quantum results lie on top of each other.}
\end{figure}
where the quantities $\kappa$ and $\Tilde{G}(\omega,t)$ have been defined in Eqs. \ref{eqn:kappa_LO_huot} and \ref{eqn:EE_finite_frequency}, respectively. $\ket{\psi_{0}}$ and $\ket{\psi_{f}}$ are the initial state and final states which are connected by a dipole-transition ($\Delta l = \pm 1$).

The production cross section of quarkonium states in heavy nuclei relative to proton-proton collisions is still an active area of research (see \cite{Brambilla:2010cs} and references therein). Different initial states have been used to calculate the survival probability of quarkonia in medium. For example in \cite{brambilla20181} initial states were chosen to be a delta function in position space in $l=0$ state. In \cite{akamatsu20181,akamatsu20191} initial states were chosen to be eigenstates of the vacuum Cornell potential. To investigate the effects of size and shape of initial wavefunction, we choose first two lowest lying eigenstates of the Coulomb and Cornell potential for our calculation.

For the eigenstates of the Coulomb potential, we used the following parameters
\begin{equation}
    \label{eqn:section4_coulomb_parameters}
    V(r) = \frac{-C_F \alpha}{r},\;
    M = 4.8{\textrm{GeV}},\;
    \alpha = 0.42. 
    \end{equation}
These value of the $\alpha$ is determined by the self consistency equation
\begin{equation}
\label{eqn:alpha_at_oneloop_pNRQCD}
 1/a_{0} = M \alpha(1/a_{0}),
\end{equation}
where $a_{0}$ is the radius of the ground state of bottomonium.
For the initial states of the Cornell potential we used the following parameters
\begin{equation}
    \label{eqn:section4_cornell_parameters}
     V(r) = \sigma{\mathrm{Min}}(r,r_0) - \frac{\alpha}{r},
     \; \alpha= 0.26,\
     \sigma = 0.21{\mathrm{GeV}}^2\;,
\end{equation}
where $r_0 = 1.2 \mathrm{fm}$ is the string breaking parameter (threshold for heavy-light meson production) as determined in \cite{bali20051}. These parameters were taken from \cite{Mocsy:2007yj} (although we use $M = 4.8\mathrm{GeV}$ whereas the mass used in \cite{mocsy20071} was $M \simeq 4.6\mathrm{GeV}$).

To implement the evolution of the temperature as the QGP medium cools down, we
use the expression for a Bjorken expanding medium. These parameters were taken
from \cite{alberico20131}. 
\begin{equation}
    \label{eqn:temp_bjorken_evolution}
    T(t) = T_0 \left( \frac{t_0}{t_0+t}\right)^{\frac{1}{3}}, T_0 =
    0.475\mathrm{GeV}, t_0 = 0.6\mathrm{fm}\;,
\end{equation}
which were also used in \cite{brambilla20181}.

The survival probability as a function of time $t$ for $1S$ and $2S$ states for
the decoherence case has been presented in Fig.
\ref{fig:final_results_decoherence}. The same results for the case of
gluo-dissociation has been presented in Fig.
\ref{fig:final_results_gluodissociation}.

\begin{enumerate}
\item{The average radii $r_{\mathrm{avg}} = \langle r \rangle$ for $1S$ states
of Coulomb and Cornell potentials are $r_{\mathrm{avg}}(\mathrm{Coulomb},1S) =
0.26\mathrm{fm} $, $r_{\mathrm{avg}}(\mathrm{Cornell}, 1S) = 0.16\mathrm{fm}$.
We find that for both the cases --- decoherence and gluo-dissociation (Fig.
\ref{fig:final_results_decoherence} and Fig.
\ref{fig:final_results_gluodissociation} respectively) --- $P(t)$ values for
Coulomb $1S$ and Cornell $1S$ initial states are not very different. Since both
wavefunctions are very narrow, it makes sense that any $\vec{r}$ dependent
medium effects are comparable for them.}
\item{ For $2S$ states the
Cornell initial wavefunctions is much narrower than Coulomb one:
$r_{\mathrm{avg}}(\mathrm{Coulomb},2S) = 0.95\mathrm{fm} $,
$r_{\mathrm{avg}}(\mathrm{Cornell},2S) = 0.29\mathrm{fm}$. The difference in
$P(t)$ between Coulomb $2S$ and Cornell $2S$ originates from the huge difference in
their average radii. However, the evolution pattern is not very intuitive.}
\item{From the Figs.~\ref{fig:final_results_decoherence} and
\ref{fig:final_results_gluodissociation}, we see that despite being a much narrower state,
$P(t)$ for Cornell $2S$ is different but of the same order of magnitude as
Coulomb $2S$. In the evolution, we have taken the potential to be screened Coulomb, which is closer in form to the
Coulomb potential. Just the difference in the evolution potential to the
potential used to calculate the eigenstate, leads to a rapid change in the
wavefunction for the Cornell $2S$ state. (This effect is very prominent in
particular for decoherence.) On the other hand, we expect decoherence and
gluo-dissociation to be more important for the initially wider Coulomb state.
The competition between these is subtle. Such large effects for $2S$ arise 
from the $\qqb$ wavefunctions becoming broad with time very quickly, and suggest that other effects that we have ignored here (in particular dissipation) could play an important role and
need to be studied further. For the eignestates of the Cornell potential it
would also be natural to evolve using a non-perturbative potential obtained from the
lattice. The calculation of non-perturbative forms for both the real and imaginary parts 
for the potential at finite temperature is an active area of
research~\cite{kaczmarek20031,rothkopf20111,Bala:2019cqu}, and we leave this
exercise for future.}
\item{Comparing the $P(t)$ for $1S$ states, for two different cases, we find
that gluo-dissociation has a much stronger effect on quarkonium decay than
decoherence.}
\item{A direct comparison of our results with the ``strong coupling'' results of
Refs.~\cite{brambilla20171,brambilla20181} is not possible as
we do not work in that regime, but the closest comparison that we can consider 
is between our gluo-dissociation results and the ``weak coupling'' results of
Refs.~\cite{brambilla20171,brambilla20181}. The main difference is that $2S$
shows substantially larger suppression than $1S$ in our calculation which is
not seen in Ref.~\cite{brambilla20171,brambilla20181}. This might
be because of the following reasons ($g$ and parameters in the
Bjorken expansion are taken to be the same)
\begin{enumerate}
\item{We have included screening in the real part of the singlet and octet
potentials and this might play an important role especially for the $2S$ states}
\item{By making a choice of the hierarchy in energy states ($V(r)\ll T$), 
Refs.~\cite{brambilla20171,brambilla20181} makes an expansion in $V/T$. This
modifies both the real and imaginary parts of the potentials. We do not make a
choice in hierarchy here.}
\item{It could also be due to a difference in the choice of the initial state.}
\end{enumerate}}
\end{enumerate}

We have presented our comparison of survival probability $P(t)$ between the
classical and quantum approach in Figs.
\ref{fig:final_results_qvsc_decoherence} and
\ref{fig:final_results_qvsc_gluodissociation} for decoherence and
gluo-dissociation respectively. We only present our results for $1S$ states as
for $2S$ states, as discussed above, additional effects might play an important role. 

We note from the Figs.\ref{fig:final_results_qvsc_decoherence} and
\ref{fig:final_results_qvsc_gluodissociation} that the two approaches give
different results for both decoherence and gluo-dissociation. For both Cornell
and Coulomb $1S$, the quantum decay probability is substantially larger than
its classical counterpart. One can understand this as follows. The wavefunction
gets wider as it evolves in
time, therefore at a later time the decay rate will be much higher than what it
was at early times in the quantum scenario.  If the medium evolution is
quasi-static (the time scale over which width becomes constant is very small
compared to other time scales) and higher order contributions are not important,
one would expect that both quantum and classical approaches should give similar 
results. 

The numerical details of our calculations are as follows. We took a lattice of
spatial extent $x\in[-2.56,\;2.56]\mathrm{fm}$ with $200$ lattice points. The
states were evolved under the stochastic Hamiltonian given in
Eq.~\ref{eqn:paper_matrix_eqn1} from $t_0$ to $t_{\mathrm{max}} = 5\mathrm{fm}$ with
time-steps of size $dt = t_{\mathrm{max}}/400$. We used $500$ instances to
perform the stochastic averaging.

\section{Conclusions and Outlook}
In this paper we studied the quantum evolution of the density matrix of $\qqb$
inside a hot quark-gluon plasma. We have used the techniques of stochastic
Hamiltonian evolution to simulate the density matrix
evolution~\cite{gisin19921}. The evolution is unitary and therefore the number
of $\qqb$ pairs is conserved. The main technical advancement in this paper
over the work done in \cite{akamatsu20181,akamatsu20191} is that the
wavefunction is three dimensional and the evolution of the complete color 
structure of the $\qqb$ pair was done.   

Our starting point was the recoilless master equation derived for a $\qqb$ pair inside a
weakly coupled medium in \cite{akamatsu20151}. The master equation describes
the process of decoherence in a regime where effects of dissipation can be 
ignored. (See Ref.~\cite{akamatsu20191} for a more quantitative estimate of
dissipation effects.)

We argued that for a medium at temperature $T$ which satisfies the relation $r
m_{D} \ll 1$, for a quarkonium state of size $r$, an expansion of the
stochastic Hamiltonian in $\vec{r}$ is justified. In
Sec.~\ref{subsec:decoherence13}, we derived a small $\vec{r}$ expanded version
of the stochastic evolution operator derived in \cite{akamatsu20151} in
recoilless limit. We tested this expansion in Appendix~\ref{appex:appendix1}
for a one dimensional colorless system for which results are available from
\cite{akamatsu20181}. For the un-expanded case our results matched the results
of \cite{akamatsu20181}. We checked that the expansion in $\vec{r}$ gives
accurate results for the lowest $3$ states of the Cornell potential. 

This expansion allows one to solve the equations for three-dimensional
wavefunctions by including transitions between different angular momentum states. $\vec{r}$
expansion also makes the generation of noise much cheaper computationally.
Finally, in the $\vec{r}$ expansion, we can relate the correlator of stochastic
noise to the momentum-diffusion coefficient $\kappa$ which can be expressed as
correlator of color electric field \cite{solana20061,akamatsu20131}. A similar
correlator was derived for the process gluo-dissociation in
\cite{brambilla20081}.

Since the hierarchy between $E_{b}, T$ and $m_D$ is not very clear for the
realizable temperatures at RHIC and LHC, a Markovian evolution of the density
matrix may not be well justified. Therefore in Sec.~\ref{sec:final_results} we
proposed a modification of the stochastic noise correlator from
Eqs.~\ref{eqn:paper_noise_correlation} to Eq.~\ref{eqn:EE_finite_frequency} to
include on-shell gluons in our calculation. The main idea was to implement a
stochastic evolution equation which gives us the same decay rate as when
calculated in pNRQCD for gluo-dissociation at leading order in $g$. A
quasi-static medium evolution was assumed to perform the calculation for the
Bjorken expanding medium.
 
Finally, in the Sec.~\ref{sec:final_results} we made a comparison of the
survival probability $P(t)$ when calculated in a classical rate-equation
approach versus in a quantum approach. Typically, most phenomenological
calculations of $R_{AA}$ for quarkonium states in literature have implemented a
rate-equation approach. We call it ``classical", since a quantum evolution of
the density matrix was not done. We found that the two approaches do not always
produce the same results for the survival probability $P(t)$. The difference
depends on the initial states chosen (we considered eigenstates of the Coulomb
potential and the Cornell potential as examples) and also the form of potential
one uses to evolve. Our main results were presented in the
Sec.~\ref{sec:final_results} where we compared the survival probability between
classical and quantum approaches separately for decoherence and
gluo-dissociation.

For the $1S$ state we found that for the choice of parameters given in
Sec.~\ref{sec:final_results}, gluo-dissociation gives a substantially larger
suppression compared to decoherence. For both the cases --- decoherence and
gluo-dissociation --- we found that for $1S$ states, the survival probability
is very similar for the two choices for the initial wavefunction. However, it should be noted
that the effects of potential change on Coulomb and Cornell states are quite
different. For both cases, we found that the $2S$ states are highly suppressed
relative to $1S$ states.

For the $1S$ state we also found that the quantum calculation shows larger suppression than the
classical calculation. Finally, the dependence of classical survival probability
on the initial wavefunction is much stronger compared to its quantum
counterpart.  

Our work can be extended in several directions. Within our framework, dissipative effects\cite{akamatsu20151} can be included.
Their effects have been in studied in a recent work \cite{akamatsu20191} for
one dimensional Abelian dynamics, and our implementation can extend it to three
dimensional wavefunctions with color dynamics. 

We would also like to put our formalism on a stronger theoretical footing. A
simple conceptual advance would involve going to higher order in in $g$ in
Eq.~\ref{eqn:EE_finite_frequency}. This introduces screening and a width for
the gluons, thereby relaxing a severe approximation in our calculation of
gluo-dissociation (Sec.~\ref{sec:pNRQCD}). Eventually it would be very useful
to derive quantum evolution equations for $\qqb$ from first
principles, only assuming $M\gg1/r\gg E_b$.
Refs.~\cite{brambilla20171,brambilla20181} have derived Lindblad equations by
making choices about the hierarchy between the scales $E_b$, $T$ and $gT$. But
since these scales are not well separated, and it would be interesting if evolution
equations can be derived without making these approximations.

Eventual connection with phenomenology would require effort in other
directions. The initial production of quarkonia from $\qqb$ and the dynamics of
quarkonia in the pre-thermalized medium, $t \lesssim 1\mathrm{fm}$ have to be
investigated to get a better understanding of the initial wavefunctions and
remove an important systematic uncertainty. To make a connection to the
observed experimental $R_{AA}$ for quarkonium states, in addition, one needs to
take as a background thermal system, a realistic three dimensional hydrodynamic
simulation.

We hope to make progress on these directions in future.

\section{Acknowledgements}
We would like to acknowledge discussions with Dibyendu Bala and Sourendu Gupta.
A.T. would like to thank Shiori Kajimoto for valuable discussions. We would, in
particular, like to acknowledge many conversations with Saumen Datta. RS would
like to thank the hospitality of INT during Program INT-19-1a and IIT Guwahati
during WHEPP 2019, where part of the work was done. We acknowledge support of
the Department of Atomic Energy, Government of India, under project no.
12-R\&D-TFR-5.02-0200.

\appendix
\section{\label{appex:appendix1} A check for small $\vec{r}$ expansion}
In this section we numerically investigate a few examples to test under what
conditions the expansion of the stochastic Hamiltonian in $\vec{r}$ is a good
approximation. 

We reproduce results Ref.~\cite{akamatsu20181} obtained without making an
expansion in $\vec{r}$, checking our implementation. We then verify that for
wavefunctions smaller than the noise correlation length, the expansion works
quite well.

Ref.~\cite{akamatsu20181} investigated two kinds of systems. First they
considered states propagating in a time independent thermal system at
temperature $T$. The initial states for this system were taken to be
eigenstates of the screened Coulomb interaction (the real part of the singlet
potential). Second, they considered states propagating in a Bjorken expanding
medium. The initial state for this system was considered to be the eigenstate
of a vacuum potential of a Cornell form.

For this section we consider parameters (coupling constants, temperatures, and
parameters in the Bjorken expansion) which are the same as those considered
in Ref.~\cite{akamatsu20181} for easy comparison. These are different from
those used to obtain our final results in Sec.~\ref{sec:final_results}.

A one dimensional version of the stochastic Eq.
(\ref{eqn:aka_reduced_density}) while ignoring the color structure was
simulated in Ref.~\cite{akamatsu20181}. The evolution of the reduced system was
carried out with the stochastic Hamiltonian,
\begin{eqnarray}
\begin{aligned} 
\label{eqn:appex_paper_stochastic_unexpanded_schrodinger}
&H(\vec{r}, t)  = -\nabla_{r}^{2} /M + V(\vec{r})+\Theta(\vec{r}, t) ,\\ 
&\Theta(\vec{r}, t) = \theta_{Q}(\vec{r} / 2, t)-\theta_{\bar{Q}}(-\vec{r} / 2, t) , \\
& \langle\langle \theta^{a}(\vec{r},t)\theta^{b}(\vec{r'},t') \rangle\rangle  = \delta^{ab}\delta(t-t') D(\vec{r} - \vec{r'}).
\end{aligned}
\end{eqnarray}{}
 and $D(\vec{r})$ was modelled as a gaussian,
\begin{eqnarray}
D(\vec{r}) = \gamma e^{-r^2/l_{\mathrm{corr}}^2}\;. 
\end{eqnarray}{}
\subsection{\label{appex:static_small_r} Time independent background}
The strength $\gamma$ was taken to be
\begin{eqnarray}
\gamma = \frac{g^2C_F T}{4\pi},
\end{eqnarray}{}
which includes the factor $C_F$ which carries the imprint of color in the
Abelian dynamics. In weak coupling $l_{\corr}~\sim gT$
(Eq.~\ref{eqn:akadrHTL}). However, since the hierarchy between $T$ and $gT$ 
($g\approx 1.7$ for $\frac{g^2C_F}{4\pi}=0.3$) is unclear,
Ref.~\cite{akamatsu20181} considered a range of values of $l_{\corr}$ varying
from $0.04$fm to $0.96$fm. Here we compare our results for two values within this
range,
\begin{eqnarray}
l_{\mathrm{corr}} \in\{\frac{1}{gT},\; \frac{1}{T}\}\;.
\end{eqnarray}{}

The stochastic fields are correlated over length $l_{\corr}$. Therefore, if the hierarchy
$l_{\corr}\gg r$ is satisfied, one can expand the stochastic fields present in
the above equations around $\vec{r} = 0$. Then up to leading order in
$r/l_{\corr}$, we expect the system to interact with the environment as a
dipole. The stochastic Schr\"{o}dinger equation in this approximation can be
written as:  

\begin{eqnarray}
\begin{aligned} 
\label{eqn:appex_paper_r_expanded_Schrodinger}
&H(\boldsymbol{r}, t)  \equiv-\nabla_{r}^{2} / (2 m)+V(\boldsymbol{r})+\boldsymbol{r}\cdot\Theta'(t) ,\\ 
&\Theta'(t) \equiv \nabla \Theta(\boldsymbol{r},t)|_{\boldsymbol{r}=0}  ,\\
& \langle\langle \Theta^{'a}(t)\Theta^{'b}(t') \rangle\rangle  = \delta^{ab}\delta(t-t') \nabla^2 D(0).
\end{aligned}
\end{eqnarray}{}

In this section we would like to examine how reliable such an expansion is in
practice. EFTs like pNRQCD and its various versions are based on these
assumptions and this exercise provides with a simple check.

Since we are in a one-dimensional system, let us make our naming scheme clear.
We label the lowest lying ground state for the given Hamiltonian by $n_1=0$.
The first and second excited states are respectively labelled as $n_1=1,\;2$.
The averaged squared radius is defined by $r_{\mathrm{rms}} 
= \sqrt{\langle r^2\rangle}$.

For the Debye screened potential used in \cite{akamatsu20181}, 
\begin{equation}
V(r) = \frac{\alpha_{\mathrm{eff}}}{r}e^{-m_D r},
\end{equation}
with parameters given in Table~\ref{tab:appex_table_static} the value of $r_{\mathrm{rms}}$ for the first two bound states is
\begin{eqnarray}
&r_{\mathrm{rms}}(n_1=0)= 0.11\mathrm{fm}\\
&r_{\mathrm{rms}}(n_1=1)= 0.66\mathrm{fm}.
\end{eqnarray}

\begin{table}[h]
\begin{ruledtabular}
\begin{tabular}{ccddd}
M[GeV]& $\alpha_{\mathrm{eff}}$&
\multicolumn{1}{c}{$m_D[\mathrm{GeV}]$}&
\multicolumn{1}{c}{$\gamma[\mathrm{GeV}]$}&
\multicolumn{1}{c}{$1/l_{\mathrm{corr}}[\mathrm{GeV}]^{-1}$}\\
\hline
4.8 & 0.3 & T & 0.3T & T,\;gT
\end{tabular}
\end{ruledtabular}
\caption{\label{tab:appex_table_static}Mass and parameters in the model used in \cite{akamatsu20181}}
\end{table}

 A convenient dimensionless quantity to characterize the separation of scales
is $\xi = r_{\textrm{rms}}/l_{\textrm{corr}}$ which we want to be much smaller
than $1$. In this section we take $T=0.4\textrm{GeV}$ to be constant in time. 

Taking $l_{\mathrm{corr}}=1/T$ gives the value $l_{\mathrm{corr}}=0.5$fm.
For $g$ corresponding to $\alpha_{\mathrm{eff}}$ in Table~\ref{tab:appex_table_static}, the
second value of $l_{\mathrm{corr}}$ is $1/gT=0.3\textrm{fm}$. 

\begin{figure}
\mbox{
\includegraphics[width=0.5\textwidth]{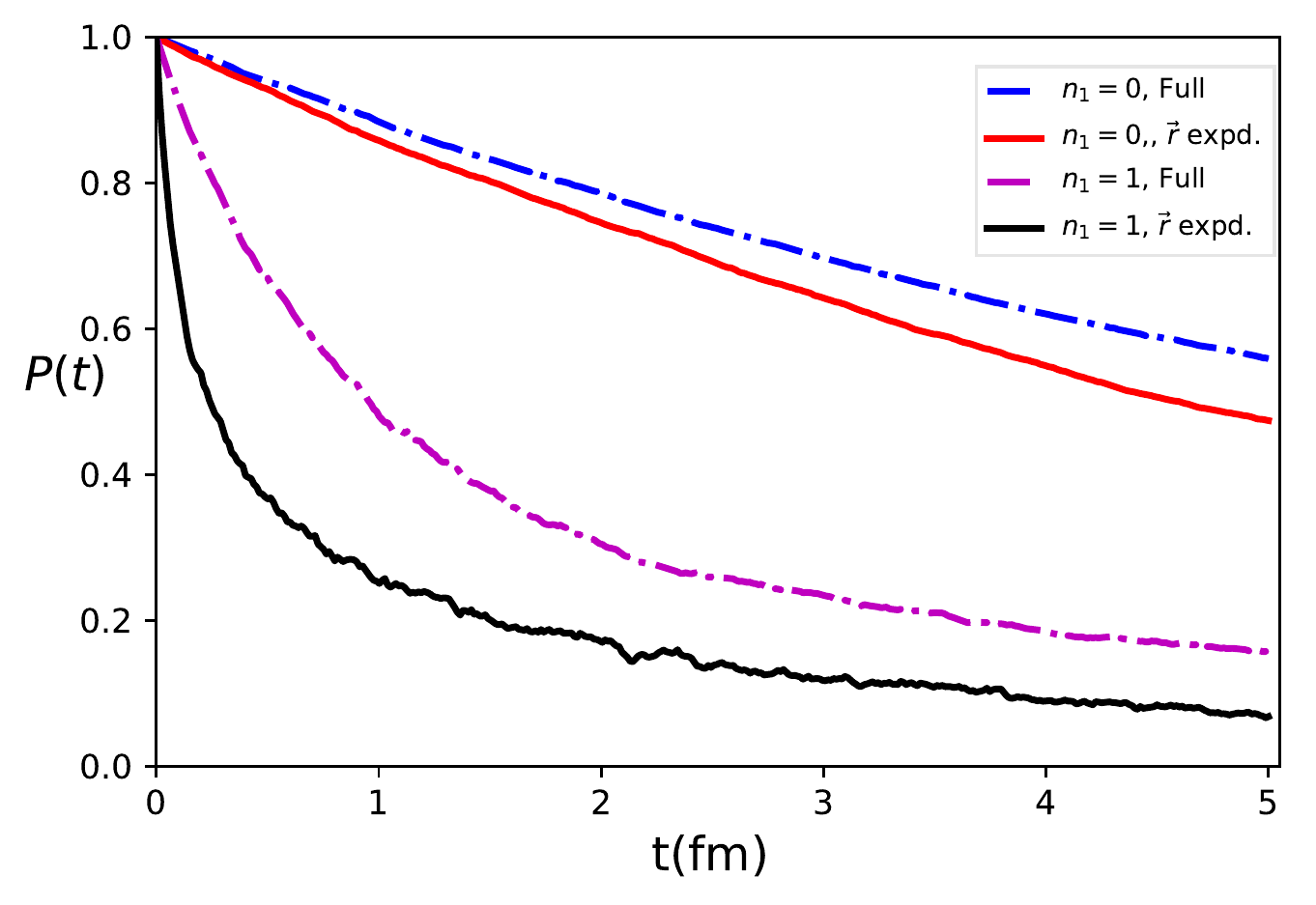}
}
\caption{\label{fig:appex_static_lcorr032}
(color online) Survival probability for expanded vs un-expanded case for
time-independent medium ($l_{\mathrm{corr}}\sim 0.3\mathrm{fm}$). For
$l_\corr=1/gT$, both $n_1=0$ (red solid and blue dash dotted lines) and $n_1=1$
(black solid and pink dash dotted lines) start differing as soon as
$t=1.5\textrm{fm}$ by more than $5\%$.}
\end{figure}

\begin{table}[h]
\begin{tabular}{c|c|c}
 & $\xi(n_{1} = 0)$ & $\xi(n_{1} = 1)$ \\
\hline
$l_{\mathrm{corr}} = 1/T$ & 0.22 & 1.32 \\
\hline
$l_{\mathrm{corr}} = 1/gT$ & 0.37 & 2.2 \\
\end{tabular}
\caption{\label{tab:appex_table_comparison_rvslcorr_debye}}
\end{table}
Taking $l_\corr=1/T$, we find that $\xi$ is indeed smaller than $1$ for the
ground state. For the $n_1=1$ state the ratio is not small and one expect that
$\vec{r}$ expansion will break down. For $l_\corr=1/(gT)$, the ratios are even larger due to $g>1$.  This expectation is indeed verified in our results.

\begin{figure}
\mbox{
\includegraphics[width = 0.5\textwidth]{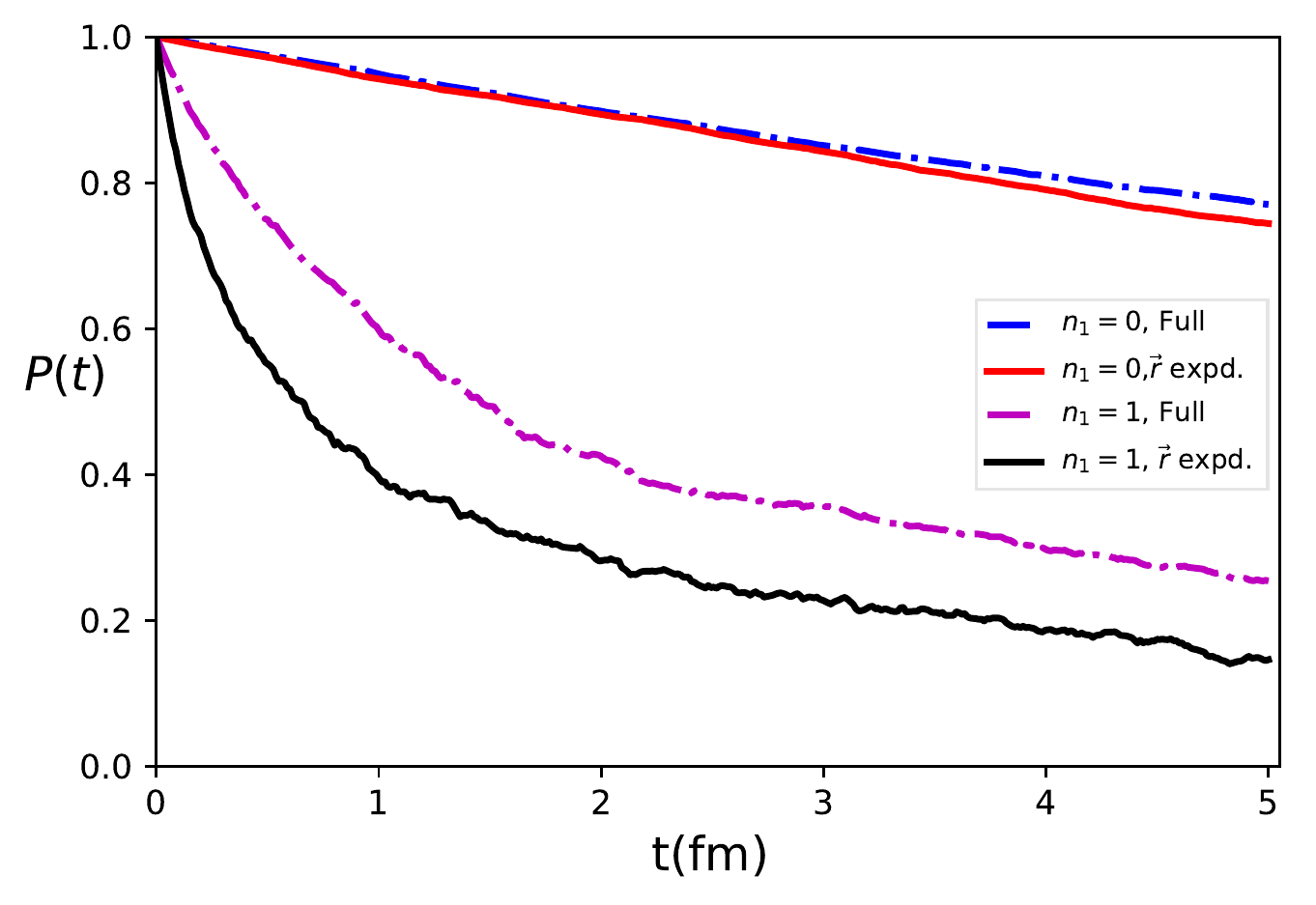}
}
\caption{\label{fig:appex_static_lcorr048}
(color online) Survival probability for expanded vs un-expanded case for
time-independent medium ($l_{\mathrm{corr}}= 0.5\mathrm{fm}$). For $n_1 =0$
state (red solid vs blue dash dotted lines) the difference is very small. For
$n_1=1$ state (black solid vs pink dash dotted lines) even at early times the
two case start differing.}
\end{figure}

The results for survival probability $P(t)$ (see Eq. ~\ref{eqn:survival_probability}) for $l_\corr=1/T$ are in given in
Fig.~\ref{fig:appex_static_lcorr048} and for $l_\corr=1/(gT)$ in
Fig.~\ref{fig:appex_static_lcorr032}. At $t = 5\textrm{fm}$, which is the stopping time for our evolution, for $l_\corr = 1/T$ we get ($P(t)$ for un-expanded and $\bar{P}(t)$ for expanded case)  
\begin{eqnarray}
\label{eqn:appex_comparison_RAA_expvsunexp_static_lcorr05}
    & P(t = 5\mathrm{fm})(n_{1} = 0) = 0.77,\quad \bar{P}(t = 5\mathrm{fm})(n_{1} = 0) = 0.74 ,
 \nonumber \\
    & P(t = 5\mathrm{fm})(n_{1} = 1) = 0.25,\quad \bar{P}(t = 5\mathrm{fm})(n_{1} = 1) = 0.15 \nonumber \\
\end{eqnarray}
and for $l_\corr = 1/gT$ ~\ref{fig:appex_static_lcorr032}
\begin{eqnarray}
    \label{eqn:appex_comparison_RAA_expvsunexp_static_lcorr03}
    & P(t = 5\mathrm{fm})(n_{1} = 0) = 0.56,\quad \bar{P}(t = 5\mathrm{fm})(n_{1} = 0) = 0.47 
, \nonumber \\
    & P(t = 5\mathrm{fm})(n_{1} = 1) = 0.16,\quad \bar{P}(t = 5\mathrm{fm})(n_{1} = 1) = 0.07 .\nonumber \\
\end{eqnarray}
We see that for $l_\corr =1/T$, the expanded vs un-expanded results are very close for $n_{1} =0$ state. For $l_\corr = 1/T$ at final time the expanded result is $3.8\%$ smaller than the un-expanded case. For $l_\corr = 1/gT$ at final time the expanded result is $16\%$ smaller than the un-expanded case.

For $n_{1}=1$ state we see that $\vec{r}$ expansion breakdowns and therefore is not reliable. For $l_\corr = 1/T$ at final time the expanded result is $40\%$ smaller than the un-expanded case. For $l_\corr = 1/gT$ at final time the expanded result is $56\%$ smaller than the un-expanded case. We conclude that for the realistic system with initial states of similar size,
$\vec{r}$ expansion is not a reliable tool.

To understand the implications for the three dimensional calculation, we note
that eigenstates in the one dimensional problem and the three dimensional
problem are related. For a rotationally invariant potential, the radial part 
of a three dimensional Schr\"{o}dinger equation is equivalent to a
one dimensional Schr\"{o}dinger equation with the additional constraint that
the wavefunction is $0$ at the origin for $l =0$ states, $l$ being the quantum number for orbital angular momentum. The $n_1=0$ state is finite at the
origin and does not correspond to any three dimensional state. The $n_1=1$
state corresponds to the $n=0$ three dimensional, $l=0$ state. Therefore our
results show that for an eigenstate of the screened Coulomb potential in
three dimensions with value of $\alpha$ similar to table \ref{tab:appex_table_static} , the $\vec{r}$ expansion is invalid and a full three
dimensional simulation is necessary. However, it is reasonable to argue that in
a rapidly evolving plasma, it is not well motivated to use the eigenstate of
the screened Coulombic state as the initial state of evolution anyway. It is more
appropriate to start the evolution with a narrow state, which is often taken in
the literature to be the state in the vacuum. This is the system we analyze in the next
section.

Finally, we comment on some technical aspects of our implementation.  We took a
lattice of spatial extent $x\in[-2.56,\;2.56]\mathrm{fm}$ with $512$ lattice
points. The states were evolved under the stochastic Hamiltonian given in
Eq. (\ref{eqn:appex_paper_stochastic_unexpanded_schrodinger} and
\ref{eqn:appex_paper_r_expanded_Schrodinger}) for $t_{\mathrm{max}} = 5\mathrm{fm}$
with time-steps of size $dt = t_{\mathrm{max}}/5000$. We used $1000$ ensembles to
perform the stochastic averaging.

\subsection{\label{appex:bjorkensmallr}Non-equilibrium QGP}

\begin{figure}
\mbox{
\includegraphics[width = 0.5\textwidth]{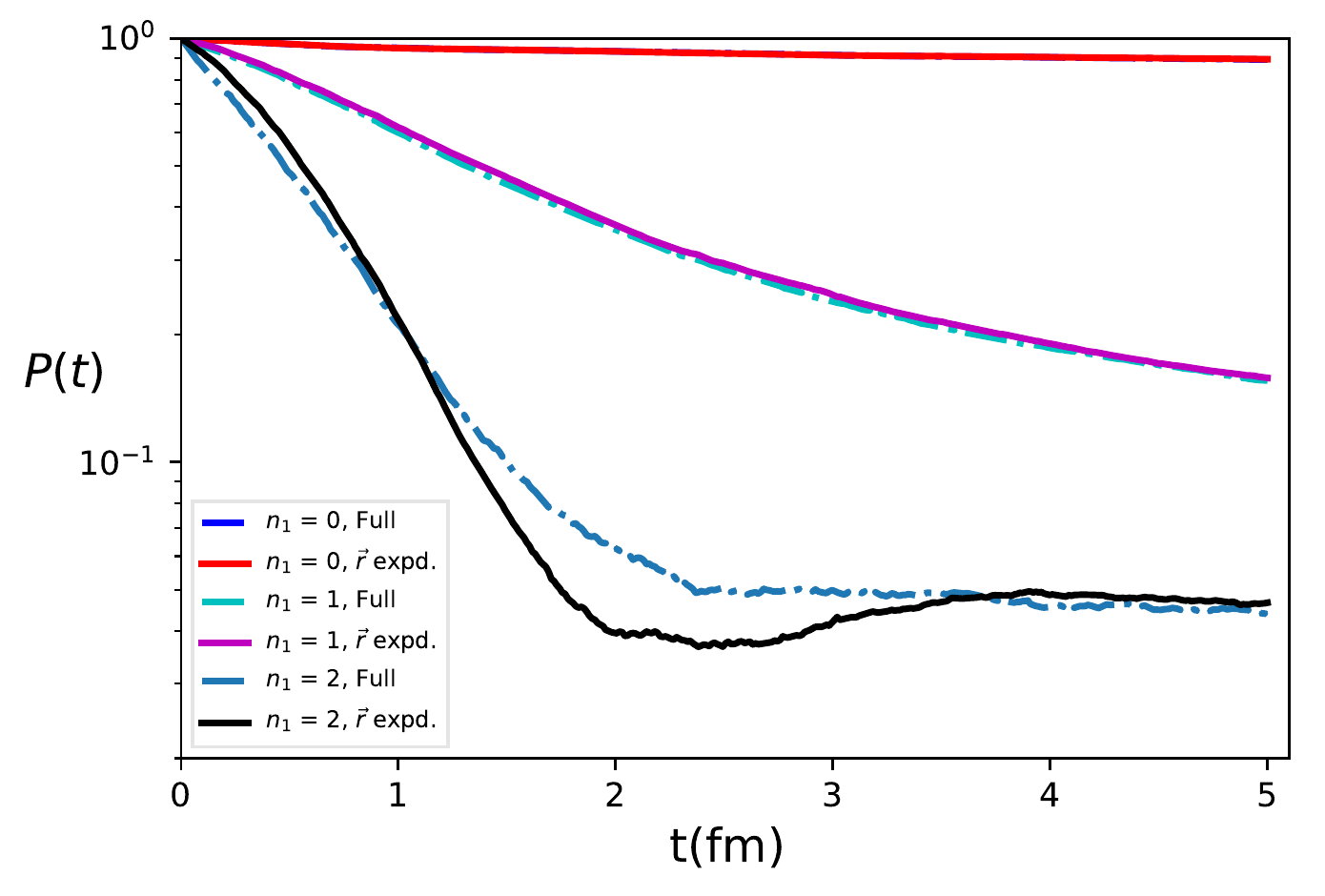}
}
\caption{\label{fig:appex_bjor_cornell} (color online) Survival probability for
a Bjorken expanding medium. For Cornell states, $n_1=0$ (red solid and blue
dash dotted lines) and $n_1=1$ (pink solid and cyan dash dotted lines) expanded
vs un-expanded results are very close to each other. Only for $n_1=2$ state
(black solid and sky-blue dash dotted lines), we see a big difference around
$t\simeq 2\textrm{fm}$ between the two cases, which narrows down as the
wavefunctions evolves further in time. }
\label{fig:bjorcomp}
\end{figure}

The Bjorken expansion is a well studied model for QGP dynamics and in this
section we present the same comparison in the case of a Bjorken expanding
medium. The temperature changes with time according to the relation,  
\begin{equation}
T(t) = T_{0}\left( \frac{t_{0}}{t+t_{0}}\right)^{(1/3)}.
\end{equation}
Although the full dynamics of the evolution of medium can be quite
complicated, for a simple calculation the picture of the Bjorken expansion is a
good check. The initial temperature was chosen to be $T_{0} = 0.4$ and $t_{0}
= 1\textrm{fm}$ to match Ref.~\cite{akamatsu20181}. 

The initial states were chosen as the first three bound states
of the one dimensional Cornell potential,
\begin{equation}
V_{\mathrm{Cornell}}(r) = \sigma r - \frac{\alpha_{\mathrm{eff}}}{r},
\sigma = 0.2\mathrm{GeV},\alpha_{\mathrm{eff}} = 0.3
\end{equation}
and the parameters of the model are same as in Table~\ref{tab:appex_table_static}.

After solving for the eigenstates the Cornell potential in one dimension, we obtain the following
values for $r_{\mathrm{rms}}$
\begin{eqnarray}
\label{eqn:appex_radius_rms_cornell}
&& r_{\mathrm{rms}}(n_1 = 0) = 0.098\mathrm{fm} \nonumber \\
&& r_{\mathrm{rms}}(n_1 = 1) = 0.27\mathrm{fm} \nonumber \\
&& r_{\mathrm{rms}}(n_1 = 2) = 0.53\mathrm{fm} 
\end{eqnarray}{}
In this section we only show the results for $l_{\corr}=1/T$. As both the
wavefunction width and the temperature change with time in this system, it is
not convenient to quote a single number $\xi$ to check whether the expansion in
$\vec{r}$ will be accurate. We simply compare the results with and without the
approximation to see how well it works. The result is presented in the Fig.
\ref{fig:appex_bjor_cornell}. At $t = 5\textrm{fm}$ the difference between the expanded and un-expanded cases is given below,

\begin{eqnarray}
    \label{eqn:appex_comparison_RAA_expvsunexp_cornell}
    & P(t = 5\mathrm{fm})(n_{1} = 0) = 0.89,\quad \bar{P}(t = 5\mathrm{fm})(n_{1} = 0) = 0.89 ,
 \nonumber \\
    & P(t = 5\mathrm{fm})(n_{1} = 1) = 0.16,\quad \bar{P}(t = 5\mathrm{fm})(n_{1} = 1) = 0.16,
 \nonumber \\
    & P(t = 5\mathrm{fm})(n_{1} = 1) = 0.047,\quad \bar{P}(t = 5\mathrm{fm})(n_{1} = 1) = 0.044 \nonumber \\
\end{eqnarray}

For the Bjorken case we find that the $\vec{r}$
expansion is a much better approximations. The largest different between the expanded vs un-expanded case is for $n_1=2$ case, which at $t= 5\textrm{fm}$ is $6\%$. This originates from two facts 

\begin{enumerate}
\item{The Cornell wavefunctions are much narrower compared to the Debye
screened potential ones. Even for the $n=2$ wavefunction, the value of $\xi
\sim 1$ at the earliest time.}
\item 
 As the medium cools down, the $l_{\mathrm{corr}}$
grows larger. At late times, it will be harder for the medium to resolve
the details of $\qqb$. For e.g. at $t=0$, the typical correlation lengths
are of the order $l_{\textrm{corr}}=0.5\textrm{fm}$ which grows up
to $l_{\textrm{corr}}=0.9\textrm{fm}$ by $t = 5\textrm{fm}$.
\end{enumerate}

Translating this to three dimensions, the $n_1=1$ and the $n_1=2$ states correspond to $1S$ and $2S$ states of a realistic three dimensional system. We conclude that at least for the initial states chosen from the Cornell like
potential evolving in a Bjorken expanding medium, $\vec{r}$ expansion can be used reliably. 

Similarly, for the lowest three eigenstates of the three dimensional 
Coulomb potential with larger value of $\alpha_{\mathrm{eff}}$,
such as those used in \cite{brambilla20181}, one can use the 
$\vec{r}$ expansion to simplify the calculation.

\bibliography{apssamp}

\end{document}